\documentclass[11pt]{article}
\usepackage[left=2.5cm, right=2cm, top=2.50cm, bottom=2cm]{geometry}
\usepackage{mathtools, mathpazo, subfigure, float, booktabs, tensor, multirow, cite}
\usepackage{makecell}
\usepackage{amsthm,amsmath,amssymb}
\usepackage{listings}
\usepackage{xcolor, color}
\usepackage{tikz}
\usepackage{graphicx}
\usepackage{epstopdf}
\usepackage{framed}
\usepackage{hyperref}
\usepackage{threeparttable}
\usepackage{booktabs}
\usepackage{comment}

\definecolor{melon}{RGB}{230,57,70}

\newcommand\mb[1]{\mathbb{#1}}
\newcommand\mr[1]{\mathrm{#1}}

\newcommand\R{\mb{R}}

\newcommand\Q{\mb{Q}}

\newcommand\de{\mr{d}}

\newcommand\br[1]{\left(#1\right)}
\newcommand\fa[1]{\left[#1\right]}
\newtheorem{thm}{Theorem}[section]

\newtheorem{lem}{Lemma}[section]
\newtheorem{proposition}{Proposition}[section]

\theoremstyle{remark}
\newtheorem{rem}{Remark}

\theoremstyle{definition}

\allowdisplaybreaks

\DeclareMathAlphabet{\mathcal}{OMS}{cmsy}{m}{n}
\DeclareSymbolFont{largesymbols}{OMX}{cmex}{m}{n}
\def\lc{\left\lceil}
\def\rc{\right\rceil}
\title{Neural Option Pricing for Rough Bergomi Model}
\author{Changqing Teng \thanks{Department of Mathematics, The University of Hong Kong, Pokfulam Road, Hong Kong. Email: {\tt{u3553440@connect.hku.hk}}.} \quad and \quad Guanglian Li\thanks{Department of Mathematics, The University of Hong Kong, Pokfulam Road, Hong Kong. Email: {\tt{lotusli@maths.hku.hk}} GL acknowledges the support from GRF (project number: 17317122) and Early Career Scheme (Project number: 27301921), RGC, Hong Kong.}}
\begin{document}
\maketitle
\begin{abstract}
The rough Bergomi (rBergomi) model can accurately describe the historical and implied volatilities, and has gained much attention in the past few years. However, there are many hidden unknown parameters or even functions in the model.
In this work we investigate the potential of learning the forward variance curve in the rBergomi model using a neural SDE.
To construct an efficient solver for the neural SDE, we propose a novel numerical scheme for simulating the volatility process using the modified summation of exponentials. Using the Wasserstein 1-distance to define the loss function, we show that the learned forward variance curve is capable of calibrating the price process of the underlying asset and the price of the European style options simultaneously. Several numerical tests are provided to demonstrate its performance.
\end{abstract}

\section{Introduction}
Empirical studies of a wild range of assets volatility time-series show that the log-volatility in practice behaves similarly to fractional Brownian motion (FBM) (FBM with a Hurst index $H$ is the unique centered Gaussian process $W^H(t)$ with autocorrelation being  $\mathbb{E}[W^H_t W^H_s]=\tfrac{1}{2}[|t|^{2H}+|s|^{2H}-|t-s|^{2H}]$) with a Hurst index $H \approx 0.1$ at any reasonable time scale \cite{gatheral2022volatility}. Motivated by this empirical observation, several rough stochastic volatility models have been proposed, all of which are essentially based on fBM and involve the fractional kernel. The rough Bergomi (rBergomi) model \cite{bayer2016pricing} is one recent rough volatility model that has a remarkable capability of fitting both historical and implied volatilities.

The rBergomi model can be formulated as follows.  Let $S_t$ be the price of underlying asset with the time horizon $[0, T]$ defined on a given filtered probability space $(\Omega, \mathcal{F}, \{\mathcal{F}_t\}_{t \geq 0}, \Q)$, with $\Q$ being a risk-neutral martingale measure:
\begin{equation}\label{eq: rBergomi}
S_t = S_0\exp\br{-\frac{1}{2}\int_{0}^tV_s\de s + \int_{0}^t\sqrt{V_s}\de Z_s},\quad t\in [0, T].
\end{equation}
Here, $V_t$ is the spot variance process satisfying
\begin{equation}\label{eq: spot variance}
V_t = \xi_0(t)\exp\br{\eta\sqrt{2H} \int_{0}^{t}(t-s)^{H - \frac{1}{2}}\de W_s- \frac{\eta^2}{2}t^{2H}},\quad t\in[0, T].
\end{equation}
$S_0>0$ denotes the initial value of the underlying asset, and the parameter $\eta$ is defined by
\begin{align*}
\eta:=2\nu \sqrt{\frac{\Gamma(3/2-H)}{\Gamma(H+1/2)\Gamma(2-2H)}},
\end{align*}
with $\nu$ being the ratio between the increment of $\log V_t$ and the FBM over $(t,t+\Delta t)$ \cite[Equation (2.1)]{bayer2016pricing}.

The Hurst index $H \in (0, {1}/{2})$ reflects the regularity of the volatility $V_t$. $\xi_0(\cdot)$ is the so-called initial forward variance curve, defined by $\xi_0(t) = \mathbb{E}^{\mathbb{Q}}[V_t | \mathcal{F}_0]= \mathbb{E}[V_t]$ \cite{bayer2016pricing}. $Z_t$ is a standard Brownian motion:
\begin{equation}
Z_t := \rho W_t + \sqrt{1-\rho^2}W^\perp_t  \label{eq: BM for S}.
\end{equation}
Here, $\rho \in (-1, 0)$ is the correlation parameter and $W^\perp_t$ is a standard Brownian motion independent of $W_t$.
The price of a European option with payoff function $h(\cdot)$ and expiry $T$ is given by
\begin{align}\label{eq:martingale}
P_0=\mathbb{E}[\mathrm{e}^{-rT}h(S_T)].
\end{align}
Despite its significance from a modeling perspective, using a singular kernel in the rBergomi model leads to the loss of Markovian and semimartingale structure. In practice, simulating and pricing options using this model involves several challenges. The primary difficulty arises from the singularity of the fractional kernel
\begin{align}\label{eq: fractional kernel}
G(t) := t^{H - \frac{1}{2}},
\end{align}
at $ t=0$, which impacts the simulation of the Volterra process given by:
\begin{equation}
I_t := \sqrt{2H}\int_0^t G(t-s)\de W_s. \label{eq: stochastic volterra process}
\end{equation}
Note that this Volterra process is a Riemann-Liouville FBM or L\'{e}vy's definition of FBM up to a multiplicative constant \cite{mandelbrot1968fractional}. The deterministic nature of the kernel $G(\cdot)$ implies that $I_t$ is a centered, locally $(H-\epsilon)$-H\"{o}lder continuous Gaussian process with $\mathbb{E}[I_t^2]=t^{2H}$. Furthermore, a straightforward calculation shows
\begin{align*}
\mathbb{E}[I_{t_1}I_{t_2}]=t_1^{2H}C\left(\frac{t_2}{t_1}\right),\quad  \text{ for } t_2>t_1,
\end{align*}
where for $x>1$, the univariate function $C(\cdot)$ is given by
\begin{align*}
C(x):=2H\int_{0}^{1}(1-s)^{H-1/2}(x-s)^{H-1/2}\de s,
\end{align*}
indicating that $I_t$ is non-stationary.

Even though the rBergomi model enjoys remarkable calibration capability with the market data, there are many hidden unknown parameters even functions. For example, the Hurst exponent (in a general SDE) is regarded as an unknown function $H := H(t)\colon \R_{+} \to (0, 1)$ parameterized by a neural network in \cite{tong2022learning}, which is learned using neural SDEs. According to \cite{bayer2016pricing}, $\xi_0(t)$ can be any given initial forward variance swap curve consistent with the market price. In view of this fact and the high expressivity of neural SDEs \cite{tzen2019neural,tzen2019theoretical, li2020scalable, kidger2021efficient, liu2019neural, jia2019neural, hodgkinson2020stochastic}, in this work, we propose to parameterize the initial forward variance curve $\xi_0(t)$ using a feedforward neural network:
\begin{align}\label{eq:neural-variance}
\xi_0(t)=\xi_0(t;\theta),
\end{align}
where $\theta$ represents the weight parameter vector of the neural network. The training data are generated via a suitable numerical scheme for a given target initial forward variance curve $\xi_0(t)$.

Motivated by the implementation in \cite{delise2021neural}, we adopt the Wasserstein metric as the loss function to train the weights $\theta$. Since the Wasserstein loss is convex and has a unique global minima, any SDE maybe learnt in the infinite data limit \cite{kidger2021neural}. Remarkably, the attained Wasserstein-1 distance during the training is a natural upper bound for the discrepancy between the exact option price $P_0$ and the learned option price $P_0(\theta^*)$ given by the neural SDE. This in particular implies that if the approximation of the underlying dynamics of stock price $S_t$ is accurate to some level, then  the European option price $P_0$ with this stock as the underlying asset will be accurate to the same level. In this manner, the loss can realize two optimization goals simultaneously.

To generate the training data to learn $\theta$, we need to jointly simulate the Volterra process $I_t$ and the underlying Brownian motion $Z_t$ for the stock price. The non-stationarity feature and the joint simulation requirement make Cholesky factorization \cite{bayer2016pricing} the only available exact method. However, it has an $\mathcal{O}(n^3)$ complexity for the Cholesky factorization, $\mathcal{O}(n^2)$ complexity and $\mathcal{O}(n^2)$ storage with $n$ being the total number of time steps. Clearly, the method is infeasible for large $n$. The most well known method to reduce the computational complexity and storage cost is the hybrid scheme \cite{bennedsen2017hybrid}.
The hybrid scheme approximates the kernel $G(\cdot)$ by a power function near zero and by a step function elsewhere, which yields an approximation combining the Wiener integrals of the power function with a Riemann sum. Since the fractional kernel $G(\cdot)$ is a power function, it is exact near zero. Its computational complexity is $\mathcal{O}(n\log n)$ and storage cost is $\mathcal{O}(n)$.

In this work, we propose an efficient modified summation of exponentials (mSOE) based method \eqref{eq: SOE for rbergomi} to simulate \eqref{eq: rBergomi}-\eqref{eq: spot variance} (to facilitate the training of the neural SDE). Similar approximations have already been considered, e.g. by Bayer and Breneis \cite{bayer2023markovian}, and Abi Jaber and El Euch \cite{abi2019multifactor}. We further enhance the numerical performance by keeping the kernel exact near the singularity, which achieves high accuracy with the fewest number of summation terms.  Numerical experiments show that the mSOE scheme considerably improve the prevalent hybrid scheme \cite{bennedsen2017hybrid} in terms of accuracy, while having less computational and storage costs. Besides the rBergomi model, the proposed approach is applicable to a wide class of stochastic volatility models, especially the rough volatility models with completely monotone kernels.

In sum, the contributions of this work are three-fold. First, we derive an efficient modified sum-of-exponentials (mSOE) based method \eqref{eq: SOE for rbergomi} to solve \eqref{eq: rBergomi}-\eqref{eq: spot variance} even with very small Hurst parameter $H\in (0,1/2)$. Second,
we propose to learn the forward variance curve by the neural SDE using the loss function based on the Wasserstein 1-distance, which can learn the underlying dynamics of the stock price as well as the European option price.
Third and last, the mSOE scheme is further utilized to obtain the training data to train our proposed neural SDE model and serves as the solver for the neural SDE, which improves the efficiency significantly.

The remaining of the paper is organized as follows. In Section \ref{sec: rBergomi simulation}, we introduce an approximation of the singular kernel $G(\cdot)$ by the mSOE, and then describe two approaches for obtaining them. Based on the mSOE, we propose a numerical scheme for simulating the rBergomi model \eqref{eq: rBergomi}-\eqref{eq: spot variance}.
We introduce in Section \ref{sec: neural option pricing} the neural SDE model and describe the training of the model. We illustrate in Section \ref{sec: experiment} the numerical performance of the mSOE scheme. Moreover, several numerical experiments on the performance of our neural SDE model are shown for different target initial forward variance curves. Finally, we summarize the main findings in Section \ref{sec: conclusion}.

\section{Simulation of the rBergomi model}\label{sec: rBergomi simulation}
In this section, we develop an mSOE scheme for efficiently simulating the rBergomi model \eqref{eq: rBergomi}-\eqref{eq: spot variance}.
Throughout, we consider an equidistant temporal grid $0 = t_0 < t_1 < \cdots < t_n = T$ with a time stepping size $\tau := {T}/{n}$ and $t_i := i\tau$.

\subsection{Modified SOE based numerical scheme}
The non-Markovian nature of the Gaussian process $I_t$ in \eqref{eq: stochastic volterra process} poses multiple theoretical and numerical challenges. A tractable and flexible Markovian approximation is highly desirable. By the well-known Bernstein's theorem \cite{Widder:1941}, a completely monotone function (i.e., functions that satisfy $(-1)^kg^{(k)}(x) \geq 0$ for all $x >0$ and $k = 0,1,2,\cdots$) can be represented as the Laplace transform of a nonnegative measure. We apply this result to the fractional kernel $G(\cdot)$ and obtain
\begin{equation}
    G(t) = \frac{1}{\Gamma(\frac{1}{2}-H)}\int_0^\infty \mathrm{e}^{-xt}x^{-H-\frac{1}{2}}\de x=\colon\int_0^\infty \mathrm{e}^{-xt}\mu(\de x) \label{eq: Berstein thm}
\end{equation}
where $\mu(\de x) = w(x)\de x$, with $w(x) = \frac{1}{\Gamma(\frac{1}{2}-H)}x^{-H-\frac{1}{2}}$. $\Gamma(\cdot)$ denotes Euler's gamma function, defined by $\Gamma(z)=\int_0^\infty s^{z-1}\mathrm{e}^{-s}\de s$ for $\Re(z)>0$. That is, $G(t)$ is an infinite mixture of exponentials.

The stochastic Fubini theorem implies
\begin{align}
    I_t &= \sqrt{2H}\int_0^tG(t-s)\de W_s = \sqrt{2H}\int_0^\infty\int_0^t \mathrm{e}^{-x(t-s)}\de W_s \mu(\de x)\nonumber\\
    &=:\sqrt{2H}\int_0^\infty Y_t^x \mu(\de x).\label{eq:xxxxx}
\end{align}
Note that for any fixed $x\geq 0$, $(Y_t^x; t\geq 0)$ is an Ornstein-Uhlenbeck process with parameter $x$, solving the SDE $\de Y_t^{x} = -x Y_t^{x}\de t + \de W_t$ starting from the origin.
Therefore, $I(t)$ is a linear functional of the infinite-dimensional process $\mathcal{Y}_t := (Y_t^x, x\geq 0)$ \cite{coutin1998fractional}. We propose to simulate $I_t$ exactly near $t$ and apply numerical quadrature to \eqref{eq: Berstein thm} elsewhere to enhance the computational efficiency, and refer the resulting method to as the modified SOE (mSOE) scheme.

Summation of exponentials can be utilized to approximate the completely monotone functions $g(x)$. This assumption covers most non-negative, non-increasing and smooth functions and is less restrictive than the requirement on the hybrid scheme \cite{romer2022hybrid}.
\begin{rem}
For the truncated Brownian semistationary process $Y_t$ of the form
$Y_t = \int_0^t g(t-s)\sigma_s\de W_s$,
the hybrid scheme requires that the kernel $g(\cdot)$ to satisfy the following two conditions:
(a) There exists an $L_g(x) \in C^1((0, 1])$ satisfying $\lim\limits_{x \to 0} \frac{L_g(tx)}{L_g(x)} = 1$ for any $t > 0$ and  the derivative $L^\prime_g$ satisfying $|{L'_g(x)}| \leq C(1 + x^{-1})$ for $x\in (0, 1]$ such that
\begin{equation*}
g(x) = x^\alpha L_g(x),\quad x\in(0,1],\ \alpha \in (-0.5, 0.5)\backslash\{0\}.
\end{equation*}
(b) $g$ is differentiable on $(0, \infty)$.
\end{rem}

Condition (a) implies that $g(\cdot)$ does not allow for strong singularity near the origin such that it can be well approximated by a certain power function. However, the rough fractional kernel $cx^{\alpha}$ with $\alpha \in [-1, -0.5]$ fails to satisfy this assumption.

Common examples of completely monotone functions are the exponential kernel $c\mr{e}^{-\lambda x}, c, \lambda \geq 0$, the rough fractional kernel $cx^{\alpha}$ with $c\geq 0$ and $\alpha<0$ and the shifted power law-kernel $(1 + t )^{\beta}$ with $\beta \leq 0$. More flexible kernels can be constructed from these building blocks as complete monotonicity is preserved by summation and products \cite[Theorem 1]{miller2001completely}.
The approximation property of the summation of exponentials is guaranteed by the following theorem.
\begin{thm}[{\cite[Theorem 3.4]{braess2012nonlinear}}]
Let $g(\cdot)$ be completely monotone and analytic for $\Re(x) > 0$, and let $0 < a < b$. Then there exists a uniform approximation $\hat{g}(x) := \sum_{j=1}^n\omega_j\mathrm{e}^{-\lambda_jx}$ on the interval $[a, b]$ such that
\begin{equation*}
    \lim\limits_{n\to\infty}\|g - \hat{g}\|_{\infty}^{1/n} \leq \sigma^{-2}.
\end{equation*}
Here, $\sigma = \exp(\frac{\pi\mathcal{K}(k)}{\mathcal{K}^\prime(k)})$, $\mathcal{K}(k) = \mathcal{K}^\prime(k^\prime)$ and $\mathcal{K}^\prime(k) = \int_{0}^\infty \frac{\de t}{\sqrt{(k^2 + t^2)(1^2 + t^2)}}$, with $k^2 + (k^\prime)^2 = 1$.
\end{thm}

The proposed mSOE scheme relies on replacing the kernel $G(\cdot)$ by $\hat{G}(\cdot)$, which is defined by
\begin{equation}
\hat{G}(t) :=\left\{
\begin{aligned}
&t^{H - \frac{1}{2}},  &&t \in [t_0, t_1),\\
&\sum_{j =1}^{N}\omega_j\mathrm{e}^{-\lambda_jt}, && t \in [t_1, t_n],
\end{aligned}
\right.  \label{eq: SOE for G}
\end{equation}
for certain non-negative paired sequence $\{(\omega_j,\lambda_j)\}_{j=1}^{N}$, where $\lambda_j$'s are the interpolation points (nodes) and $\omega_j$'s are the corresponding weights. This scheme is also referred to as the mSOE-$N$ scheme below. By replacing $G(\cdot)$ with $\hat{G}$ in \eqref{eq: fractional kernel}, we derive the associated approximation $\hat I(t_i)$ for $I_{t_i}$,
\begin{equation}\label{eq: SOE for I}
\hat{I}(t_i) := \sqrt{2H}\int_{t_{i-1}}^{t_i}(t_i -s)^{H -\frac{1}{2}}\de W_s + \sqrt{2H}\sum_{j=1}^{N} \omega_j\int_0^{t_{i-1}}\mathrm{e}^{-\lambda_j(t_i-s)}\de W_s.
\end{equation}
Then the It\^o isometry implies
\begin{align}
    \mathbb{E}\left[|{I(t_i) - {\hat{I}}(t_i)}|^2\right] &= 2H\int_{0}^{t_{i-1}}|{G(t_i-s) - \hat{G}(t_i-s)}|^2\de s\nonumber\\
    &= 2H\int_{t_1}^{t_{i}}|{G(s) - \hat{G}(s)}|^2\de s. \label{eq:isometry-xx}
\end{align}
This indicates that we only need to choose non-negative paired sequence $\{(\omega_j,\lambda_j)\}_{j=1}^{N}$ which minimize
$\int_{t_1}^{t_{i}}|{G(s) - \hat{G}(s)}|^2\de s$ in order to obtain a good approximation of $I(t_i)$ in the sense of pointwise mean square error.
Next, we describe two known approaches to obtain the non-negative paired sequence $\{(\omega_j,\lambda_j)\}_{j=1}^{N}$ in \eqref{eq: SOE for G}. Both are essentially based on Gauss quadrature.

\subsubsection*{Summation of exponentials: approach A}
Approach A in \cite{bayer2023markovian} applies $m$-point Gauss-Jacobi quadrature with weight function $x^{-H-\frac{1}{2}}$ to $n$ geometrically spaced intervals $[\zeta_i, \zeta_{i+1}]_{i=0,\cdots, n-1}$ and uses a Riemann-type approximation on the interval $[0, \zeta_0]$. The resulting approximator $\hat{G}_{A}(t)$ with $N + 1$ number of summations is given by
\begin{equation*}
    \hat{G}_{A}(t) := \sum_{j=0}^{N} \omega_j\mathrm{e}^{-\lambda_jt}\quad t\in [t_1, t_n],
\end{equation*}
where $\lambda_0 = 0$ and $\omega_0 = \frac{1}{\Gamma(\frac{1}{2}-H)}\int_0^{\zeta_0}x^{-H-\frac{1}{2}}\de x $.
Let $N:= nm$ be the prespecified total number of nodes. The parameters $m$, $n$ and $\zeta_i$ are computed
based on a set of parameters $(\alpha, \beta, a, b) \in (0, \infty)$ as follows.
\begin{align*}
    m &= \lc \frac{\beta}{A}\sqrt{N}\rc,\quad n = \left\lfloor \frac{A}{\beta}\sqrt{N} \right\rfloor (mn \approx N),\\
    A &= \br{\frac{1}{H} + \frac{1}{3/2-H}}^{1/2},\\
    \zeta_0 &= a\exp\br{-\frac{\alpha}{(3/2-H)A}\sqrt{N}},\quad
    \zeta_n = b\exp\br{\frac{\alpha}{HA}\sqrt{N}},\\
    \zeta_i &= \zeta_0\br{\frac{\zeta_n}{\zeta_0}}^{i/n},\quad i=0,\cdots, n.
\end{align*}
The following lemma gives the approximation error of the above Gaussian quadrature, which is a modification of \cite[lemmas 2.8 and 2.9]{bayer2023markovian}.
\begin{lem}
Let $(\omega_j)_{j=1}^{nm}$ be the weights and $(\lambda_j)_{j=1}^{nm}$ be the nodes of Gaussian quadrature on the intervals $[\zeta_i, \zeta_{i+1}]_{i=0, \cdots, n-1}$ computed on the set of parameters $(\alpha, \beta, 1, 1) $. Then the following error estimates hold
\begin{align*}
    \left| {\int_0^{\zeta_0} \mathrm{e}^{-xt}\mu(\de x) - \omega_0}\right| &\leq \frac{t}{\Gamma(\frac{1}{2}-H)(\frac{3}{2}-H)}\exp\left(-\frac{\alpha}{A}\sqrt{N}\right),  \\
    \left|{\int_{\zeta_0}^{\zeta_n} \mathrm{e}^{-xt}\mu(\de x)- \sum_{j=1}^{nm}\omega_j\mathrm{e}^{-\lambda_jt}}\right| &\leq \sqrt{\frac{5\pi^3}{18}}\frac{n\br{\mathrm{e}^{\alpha\beta}-1}t^{H-\frac{1}{2}}}{\Gamma(\frac{1}{2}-H)2^{2m+1}m^H} \exp\br{\frac{2\beta}{A}\sqrt{N}\log\br{\mathrm{e}^{\alpha\beta}-1}}, \\
    \left|{\int_{\zeta_n}^{\infty} \mathrm{e}^{-xt}\mu(\de x)}\right| &\leq \frac{1}{t\Gamma(\frac{1}{2}-H){\zeta_n}^{H + \frac{1}{2}}}\exp(-\zeta_nt).
\end{align*}
\end{lem}

\subsubsection*{Summation of exponentials: approach B}\label{subsec:Jiang-soe}
Approach B in \cite{jiang2017fast} approximates the kernel $G(t)$ efficiently on the interval $[t_1, t_n]$ with the desired precision $\epsilon > 0$. It applies $n_0$-point Gauss-Jacobi quadrature on the interval $[0, 2^{-M}]$ with the weight function $x^{-H-\frac{1}{2}}$ where $n_0 = \mathcal{O}(\log{\frac{1}{\epsilon}})$, $M = \mathcal{O}(\log T)$; $n_s$-point Gauss-Legendre quadrature on $M$ small intervals $[2^j, 2^{j+1}]$, $j = -M, \cdots, -1$, where $n_s = \mathcal{O}(\log{\frac{1}{\epsilon}})$, and $n_l$-point Gauss-Legendre quadrature on $N+1$ large intervals $[2^j, 2^
{j+1}]$, $j = 0, \cdots, N$, where $n_l = \mathcal{O}(\log\frac{1}{\epsilon} +\log \frac{1}{\tau})$, $N = \mathcal{O}(\log\log\frac{1}{\epsilon} +\log \frac{1}{\tau})$ \cite[Theorem 2.1]{jiang2017fast}. The resulting approximation $\hat{G}_{B}(t)$ reads,
\begin{equation}
     \hat{G}_{B}(t) = \sum_{k = 1}^{n_0}\omega_{0, k}\mathrm{e}^{-\lambda_{0, k}t} + \sum_{j = -M}^{-1}\sum_{k = 1}^{n_s}\omega_{j, k}\mathrm{e}^{-\lambda_{j,k}t}\lambda_{j,k}^{-H -\frac{1}{2}} + \sum_{j=0}^N\sum_{k=1}^{n_l}\omega_{j,k}\mathrm{e}^{-\lambda_{j,k}t}\lambda_{j,k}^{-H -\frac{1}{2}} \quad t \in [t_1, t_n],\label{eq: Jiang-soe}
\end{equation}
where the $\lambda_{j,k}^{-H -\frac{1}{2}}$ terms could be absorbed into the corresponding $\omega_{j,k}$s so that the approximation is in the form of \eqref{eq: SOE for G}. There holds $|G(t) - \hat{G}_{B}(t)| \leq \epsilon$ . For further optimization, a modified Prony's method is applied on the interval $(0, 1)$ and standard model reduction method is applied on $[1, 2^{N+1}]$ to reduce the number of exponentials needed.

The approximation error is given below, which is a slight modification of
\cite[Lemmas 2.2, 2.3 and 2.4]{jiang2017fast}.
\begin{lem}
Let $ a = 2^{-M}$, $p = 2^{N+1}$ and follow the settings in \eqref{eq: Jiang-soe}. Then the following error estimates hold
\begin{align*}
    &\left|{\int_0^a\mathrm{e}^{-xt}\mu(\de x) - \sum_{k=1}^{n_0}w_{0, k}\mathrm{e}^{-\lambda_{0. k}t}}\right| \leq \frac{4\sqrt{\pi}}{\Gamma(\frac{1}{2} - H)}a^{\frac{1}{2} - H}n_0^{\frac{3}{2}}\br{\frac{\mathrm{e}aT}{8(n_0-1)}}^{2n_0},\\
    &\left|{\int_a^p\mathrm{e}^{-xt}\mu(\de x) - \sum_{j = -M}^{-1}\sum_{k = 1}^{n_s}\omega_{j, k}\mathrm{e}^{-\lambda_{j,k}t}\lambda_{j,k}^{-H -\frac{1}{2}} -\sum_{j=0}^N\sum_{k=1}^{n_l}\omega_{j,k}\mathrm{e}^{-\lambda_{j,k}t}\lambda_{j,k}^{-H -\frac{1}{2}}}\right|\\
    &\quad\leq \frac{2^{\frac{3}{2}}\pi}{\Gamma(\frac{1}{2} - H)}\br{\frac{\mathrm{e}^{\frac{1}{\mathrm{e}}}}{4}}^{\max(n_s, n_l)}\frac{2^{(\frac{1}{2} - H)(N+1)} - 2^{-(\frac{1}{2} - H)M}}{2^{\frac{1}{2}-H} - 1},\\
    &\left|{\int_p^{\infty}\mathrm{e}^{-xt}\mu(\de x)}\right| \leq \frac{1}{\tau\Gamma(\frac{1}{2} - H)p^{H + \frac{1}{2}}}\mathrm{e}^{-\tau p}.
\end{align*}
\end{lem}

We shall compare Approaches A and B in Section \ref{sec: experiment}, the result shows that Approach B outperforms Approach A. We present in Table \ref{table:approachB} the parameter pairs for $H=0.07$ and $N=20$. See \cite[Section 3.4]{JinZhou:2023} for further discussions about SOE approximations for the fractional kernel $G(t)$.

\begin{table}
\centering
\begin{threeparttable}
\caption{Parameters for mSOE scheme based on approach B with $H = 0.07$, $\epsilon = 0.0008$ and $N = 20$.}
\begin{tabular}{c|cl|c|cl}
\toprule
$j$&  $\omega_j$ & $\lambda_j$ & $j$&  $\omega_j$ & $\lambda_j$\\
\midrule
1& 0.26118  & 0.47726  & 11 &   1.28121  &  1.92749\\
2&  0.19002 &  0.22777 & 12 & 1.76098  & 40.38675 \\
3 & 0.13840 &  0.108690 & 13 &  2.42043  & 84.62266\\
4 & 0.10717&  5.11098 $\times 10^{-2}$ &14 &  3.32681  & 177.31051\\
5 & 0.11366  &  1.93668  $\times 10^{-2}$ & 15 &   4.57262  & 371.52005\\
6&  0.14757  &  2.04153 $\times 10^{-3}$ & 16 &  6.28495 &   778.44877\\
7&  0.35898   & 1 &17 &  8.63850 &  1631.08960\\
8& 0.49341  &  2.09531 & 18 &  11.87339  &  3417.63439\\
9 &   0.67818  & 4.390310 &19 &  16.31967   &  7160.99521 \\
10 & 0.93214  &   9.19906 & 20 & 22.43096  &  15004.4875\\
\bottomrule
\end{tabular}
\label{table:approachB}
\end{threeparttable}
\end{table}

\subsection{Numerical method based on mSOE scheme}

Next we describe a numerical method to simulate $(S_{t_i},V_{t_i})$ for $i=1,\cdots,n$ based on the mSOE scheme \eqref{eq: SOE for I}. Recall that the integral  \eqref{eq: stochastic volterra process} can be written into a summation of the local part and the history part
\begin{equation} \label{eq: local and history}
    \begin{aligned}
        I_{t_{i+1}} &= \sqrt{2H}\int_{0}^{t_i}(t_{i+1} - s)^{H - \frac{1}{2}}\de W_s + \sqrt{2H}\int_{t_i}^{t_{i+1}}(t_{i+1} - s)^{H - \frac{1}{2}}\de W_s\\
        &=\colon I_{\mathcal{F}}(t_{i+1}) + {I_{\mathcal{N}}(t_{i+1})}.
    \end{aligned}
\end{equation}
The local part $I_{\mathcal{N}}(t_{i+1})\sim \mathcal{N}(0, \tau^{2H})$ can be simulated exactly. The history part $I_{\mathcal{F}}(t_{i+1})$ is approximated by replacing the kernel $G(\cdot)$ by the mSOE $\widehat G(t)$ (by approach B):
\begin{equation*}
    \bar{I}_{\mathcal{F}}(t_{i+1}) = \sqrt{2H}\sum_{j=1}^{N}\omega_j\int_{0}^{t_i}\mathrm{e}^{-\lambda_j(t_{i+1} - s)}\de W_s
    =\colon \sqrt{2H}\sum_{j=1}^{N}\omega_j\bar{I}_{\mathcal{F}}^j(t_{i+1}).
\end{equation*}
Then direct computation leads to
\begin{equation*}
\begin{aligned}
\bar{I}_{\mathcal{F}}^j(t_{i+1}) &= \mathrm{e}^{-\lambda_j\tau}\int_{0}^{t_i}\mathrm{e}^{-\lambda_j(t_i - s)}\de W_s\\
&= \mathrm{e}^{-\lambda_j\tau}\br{\int_{0}^{t_{i-1}}\mathrm{e}^{-\lambda_j(t_i-s)}\de W_s + \int_{t_{i-1}}^{t_i}\mathrm{e}^{-\lambda_j(t_i - s)}\de W_s}\\
&= \mathrm{e}^{-\lambda_j\tau}\br{\bar{I}_{\mathcal{F}}^j(t_{i}) + \int_{t_{i-1}}^{t_i}\mathrm{e}^{-\lambda_j(t_i - s)}\de W_s}.
\end{aligned}
\end{equation*}
Consequently, we obtain the recurrent formula for each history component
\begin{equation}\label{eq: SOE recursive}
\bar{I}_{\mathcal{F}}^j(t_{i})=
\left\{
\begin{aligned}
&0&& i=1,\\
&\mathrm{e}^{-\lambda_j\tau}\br{\bar{I}_{\mathcal{F}}^j(t_{i-1}) + \int_{t_{i-2}}^{t_{i-1}}\mathrm{e}^{-\lambda_j(t_{i-1} - s)}\de W_s}&& i>1.
\end{aligned}
\right.
\end{equation}
This, together with \eqref{eq: local and history}, implies that we need to simulate a centered $(N+2)$-dimensional Gaussian random vector at $t_i$,
\begin{equation*}
  {\Theta}_i:=\br{\Delta W_{t_{i}}, \int_{t_{i-1}}^{t_i}\mathrm{e}^{-\lambda_1(t_i - s)}\de W_s, \cdots, \int_{t_{i-1}}^{t_i}\mathrm{e}^{-\lambda_{N}(t_i - s)}\de W_s, I_{\mathcal{N}}(t_{i})}\quad \text{for}\ i = 1,\cdots, n.
\end{equation*}
Here, $\Delta W_{t_{i}}=:W_{t_i} - W_{t_{i-1}}$ denotes the increment. Note that ${\Theta}_i$ is determined by its covariance matrix $\Sigma$, which is defined
\begin{equation*}
\begin{aligned}
&\Sigma_{1,1} = \tau,\quad\Sigma_{1, l} = \Sigma_{l, 1} = \frac{1}{\lambda_{l-1}}\br{1 - \mathrm{e}^{-\lambda_{l-1}\tau}},\quad\Sigma_{k, l} = \frac{1}{\lambda_{k-1}+\lambda_{l-1}}\br{1 - \mathrm{e}^{-(\lambda_{k-1}+\lambda_{l-1})\tau}}\\
&\Sigma_{N+2, 1} = \Sigma_{1, N+2} = \frac{\sqrt{2H}}{H+1/2}\tau^{H+\frac{1}{2}}\\
&\Sigma_{N+2, l} = \Sigma_{l, N+2} = \frac{\sqrt{2H}}{\lambda_{l-1}^{H + 1/2}}\gamma(H + \tfrac{1}{2}, \lambda_{l-1}\tau)\\
&\Sigma_{N+2, N+2} = \tau^{2H}
\end{aligned}
\end{equation*}
for $k, l = 2, \cdots, N + 1$, where $\gamma(\cdot, \cdot)$ refers to the lower incomplete gamma function. We only need to inplement the Cholesky decomposition once since $\Sigma$ is independent of $i$.

Finally, we present the two-step numerical scheme for $(S_{t_{i+1}},V_{t_{i+1}})$ for $i = 0, \cdots, n-1$,
\begin{equation}\label{eq: SOE for rbergomi}
    \begin{aligned}
        \bar{S}_{t_{i+1}} &= \bar{S}_{t_i}\exp\br{\sqrt{\bar{V}_{t_{i}}}\br{\rho\Delta W_{t_{i+1}} + \sqrt{1 - \rho^2}\Delta W_{t_{i+1}}^\perp} - \frac{\tau}{2}\bar{V}_{t_i}},\\
\bar{V}_{t_{i+1}} &= \xi_0(t_{i+1})\exp\br{\eta\br{\bar{I}_{\mathcal{F}}(t_{i+1}) + I_{\mathcal{N}}(t_{i+1})}- \frac{\eta^2}{2}t_{i+1}^2},\\
\bar{I}_{\mathcal{F}}(t_{i+1}) &= \sqrt{2H}\sum_{j=1}^{N}\omega_j\bar{I}_{\mathcal{F}}^j(t_{i+1}),\\
        \bar{I}_{\mathcal{F}}^j(t_{i+1}) &= \mathrm{e}^{-\lambda_j\tau}\br{\bar{I}_{\mathcal{F}}^j(t_{i}) + \int_{t_{i-1}}^{t_i}\mathrm{e}^{-\lambda_j(t_i - s)}\de W_s}.\\
    \end{aligned}
\end{equation}
This mSOE-N based simulation scheme \eqref{eq: SOE for rbergomi} only requires $\mathcal{O}(N^3)$ offline computation complexity which accounts for the Cholesky decomposition of the covariance matrix $\Sigma$, $\mathcal{O}(Nn)$ computation complexity and $\mathcal{O}(N)$ storage.

\section{Learning the forward variance curve}\label{sec: neural option pricing}
This section is concerned with learning the forward variance curve $\xi_0(t;\theta)$ with $\theta$ being the weight parameters from the feedforward neural network \eqref{eq:neural-variance} by Neural SDE. Without loss of generality, assuming $S_0 = 1$, we can rewrite \eqref{eq: rBergomi}-\eqref{eq: spot variance} as
\begin{equation}\label{eq: rbergomi_v2}
\left\{\begin{aligned}
        S_t &=  1 + \int_0^tS_s\exp(X_s)\de Z_s,\\
        X_t &= \frac{1}{2}\log(V_t).
\end{aligned}\right.
\end{equation}
Upon parameterizing the forward variance curve by \eqref{eq:neural-variance}, the dynamics of the price process and variance process are also parameterized accordingly, which is given by the following neural SDE
\begin{equation}\label{eq: rbergomi_v-nn}
\left\{\begin{aligned}
        S_t(\theta) &=  1 + \int_0^tS_s(\theta)\exp(X_s(\theta))\de Z_s,\\
        X_t(\theta) &= \frac{1}{2}\log(V_t(\theta)), \\
        V_t(\theta) &=  \xi_0(t;\theta)\mathcal{E}(\eta I(t)).
\end{aligned}\right.
\end{equation}
where $\mathcal{E}(\cdot)$ denotes the (Wick) stochastic exponential.  
Then we propose to use the Wasserstein-1 distance as the loss function to train this neural SDE. That is, the learned weight parameters $\theta^*$ are given by
\begin{align}\label{eq:loss-train}
\theta^*:=\arg\min_{\theta}W_1\left(S_T, S_T(\theta)\right).
\end{align}
Next, we recall the Wasserstein distance. Let $(M, d)$ be a Radon space. The Wasserstein-$p$ distance for $p \in [1, \infty)$ between two probability measures $\mu$ and $\nu$ on $M$ with finite $p$-moments is defined by
\begin{equation*}
    W_p(\mu, \nu) = \br{\inf\limits_{\gamma \in \Gamma(\mu, \nu)}\mathbb{E}_{(x,y)\sim \gamma}d(x, y)^p}^{1/p},
\end{equation*}
where $\Gamma(\mu, \nu)$ is the set of all couplings of $\mu$ and $\nu$. A coupling $\gamma$ is a joint probability measure on $M \times M$ whose marginals are $\mu$ and $\nu$ on the first and second factors, respectively. If $\mu$ and $\nu$ are real-valued, then their Wasserstein distance can be simply computed by utilising their cumulative distribution functions \cite{villani2021topics}:
\begin{equation*}
    W_p(\mu, \nu) = \br{\int_0^1|F^{-1}(z) - G^{-1}(z)|^p\de z}^{1/p}. 
\end{equation*}
In particular, when $p=1$, according to the Kantorovich-Robenstein duality \cite{villani2021topics}, Wasserstein-1 distance can be represented by
\begin{equation}
    W_1(\mu, \nu) = \sup\limits_{\mathrm{Lip}(f) \leq 1}\mathbb{E}_{x\sim \mu}[f(x)] - \mathbb{E}_{y\sim \nu}[f(y)], \label{eq: W_1}
\end{equation}
where $\mathrm{Lip}(f)$ denotes the Lipschitz constant of $f$.

In the experiment, we work with the empirical distributions on $\mathbb{R}$. If $\xi$ and $\eta$ are two empirical measures with $m$ samples $X_1, \cdots, X_m$ and $Y_1, \cdots, Y_m$, then the Wasserstein-$p$ distance is  given by
\begin{equation}
    W_p(\xi, \eta) = \br{\frac{1}{m}\sum_{i = 1}^m|{X_{(i)} - Y_{(i)}}|^p}^{1/p},\label{eq: empirical W_p}
\end{equation}
where $X_{(i)}$ is the $i$th smallest value among the $m$ samples. With $S_t(\theta^*)$ being the price process of the underlying asset after training, the price $P_0(\theta^*)$ of a European option with payoff function $h(x)$ and expiry $T$ is given by
\begin{align}\label{eq:martingale-nn}
P_0(\theta^*)=\mathbb{E}[\mathrm{e}^{-rT}h(S_T(\theta^*))].
\end{align}
Since the payoff function $h(\cdot)$ is clearly Lipschitz-1 continuous, according to \eqref{eq: W_1}, Wasserstein-1 distance is a natural upper bound for the pricing error of the rBergomi model. Thus, the choice of the training loss is highly desirable.
\begin{proposition}
Set the interest rate $r = 0$, the Wasserstein-1 distance is a natural upper bound for the pricing error of the rBergomi model:
\begin{equation}
|P_0-P_0^*|=\left|\mathbb{E}\left[h(S_T)\right] - \mathbb{E}\left[h({S}_T(\theta^*))\right]\right| \leq W_1(S_T, S_T(\theta^*)). \label{eq: pricing error upper bound}
\end{equation}
\end{proposition}
\section{Numerical tests}\label{sec: experiment}
In this section, we illustrate the performance of the proposed mSOE scheme \eqref{eq: SOE for rbergomi} for simulating \eqref{eq: rBergomi}-\eqref{eq: spot variance}, and demonstrate the learning of the forward variance curve \eqref{eq:neural-variance} using neural SDE \eqref{eq: rbergomi_v-nn}.

\subsection{Numerical test for mSOE scheme \eqref{eq: SOE for rbergomi}}

\begin{figure}[hbt!]
    \centering
    \subfigtopskip = 2pt
    \subfigbottomskip = -2pt
    \subfigcapskip = -4pt
    \subfigure[mSOE approximation error]{
    \includegraphics[height=6cm,trim={12cm 0.3cm 12cm 2.5cm},clip]{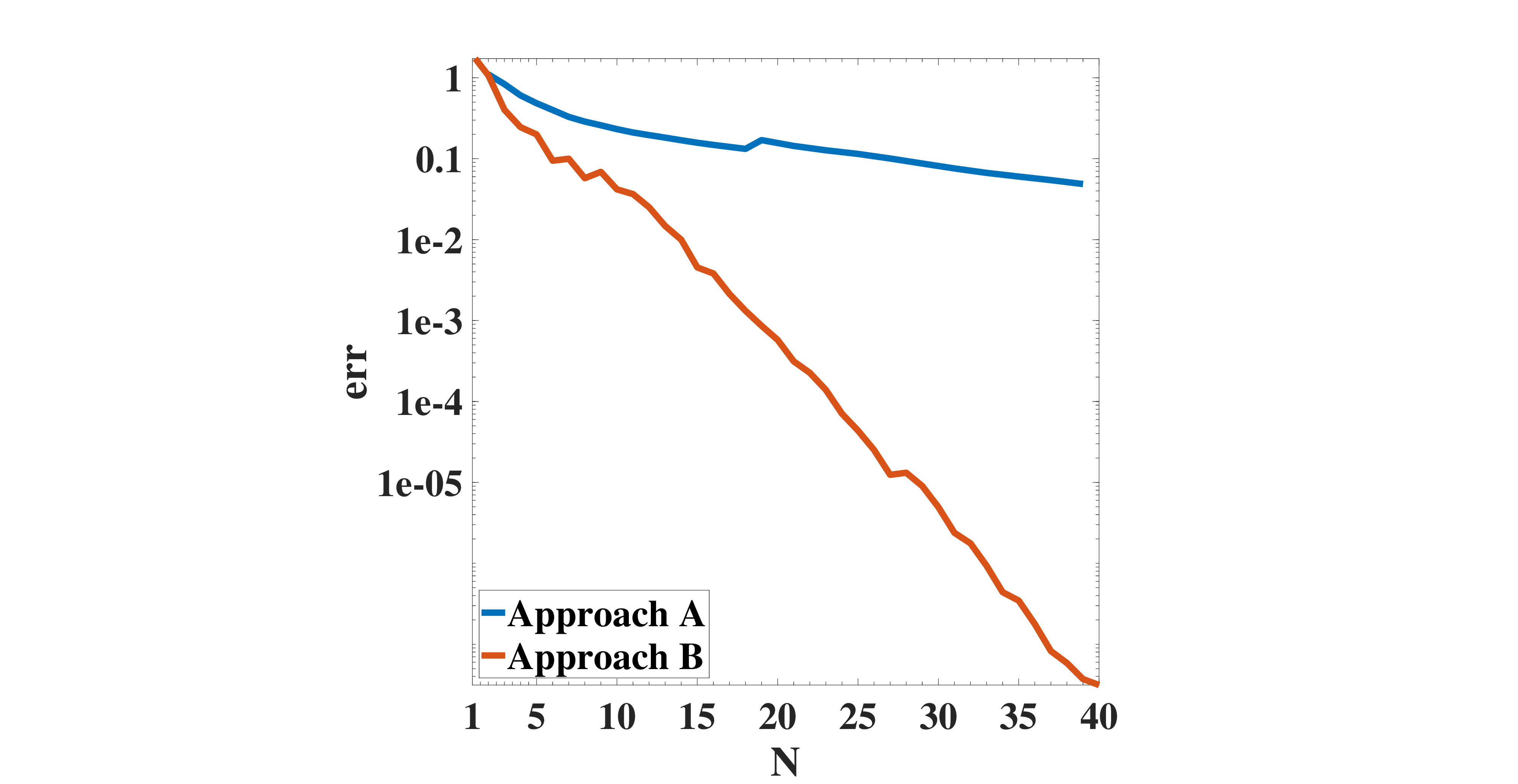}}
    \subfigure[training dynamics]{
    \includegraphics[height=6cm,trim={12cm 1cm 12cm 1cm},clip]{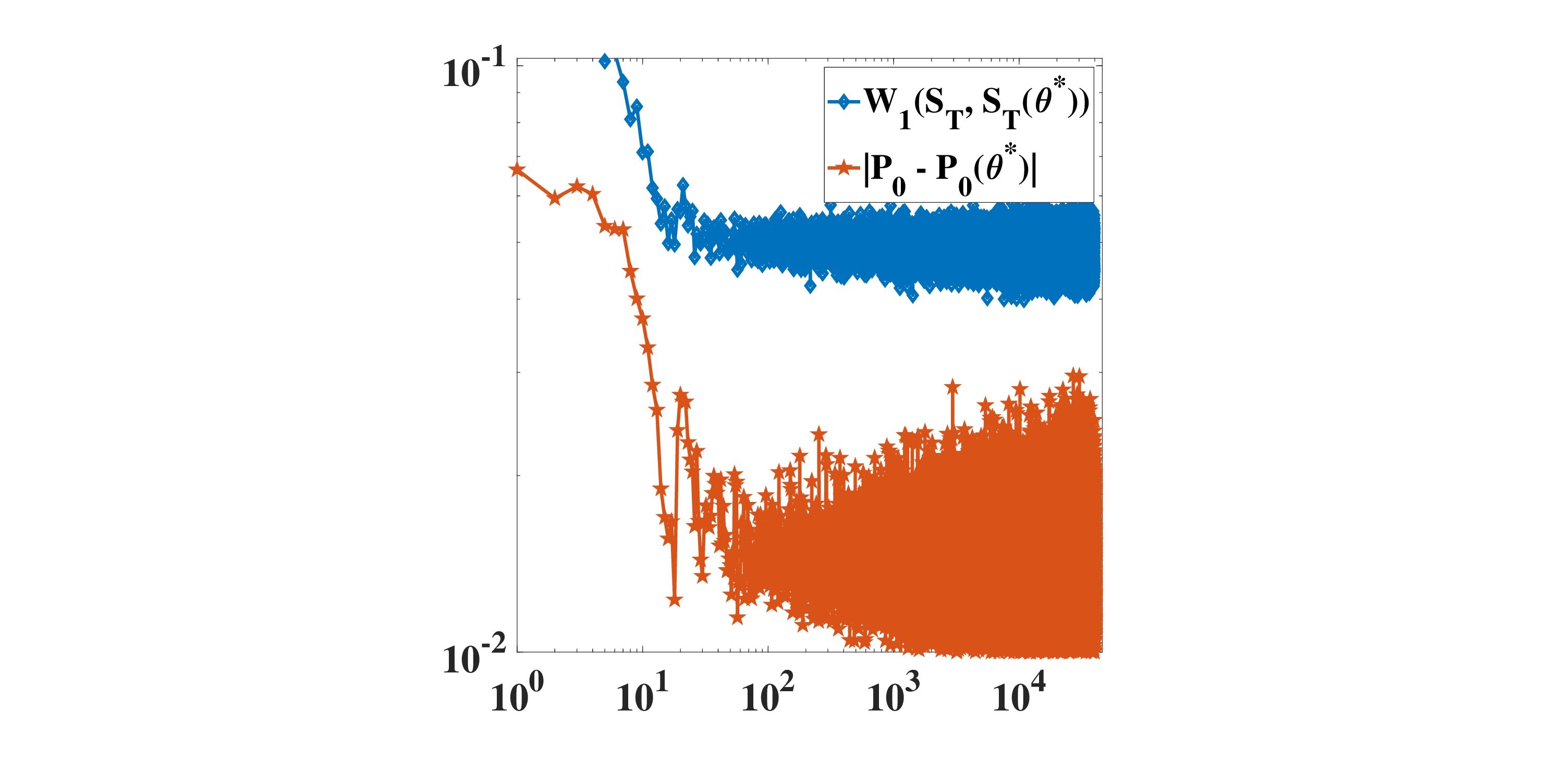}}
    \caption{(a) The performance of the mSOE-$N$ schemes based on approaches A and B with $\tau = 0.0005$, $T = 1$ and $H = 0.07$, and (b) the training dynamics.}
    \label{fig: kernel approximation}
\end{figure}

First we compare approaches A and B. They only differ in the choice of nodes $\lambda_j$ and weights $\omega_j$. To compare their performance in the simulation of the stochastic Volterra process $I(t)$, one only needs to estimate the discrepancy between $G$ and $\hat{G}$ in $L^2(t_1,T)$-norm \eqref{eq:isometry-xx}. Thus, as the criterion for comparison, we use $\text{err} := (\int_{t_1}^T(G(t) - \hat{G}(t))^2\de t)^{\frac{1}{2}}$.
We show in Figure \ref{fig: kernel approximation}(a) the $\text{err}$ versus the number of nodes $N$ for approaches A and B with $H=0.07$. We observe that approach B has a better error decay. This behavior is also observed for other $H\in (0,1/2)$. Thus, we implement approach B in \eqref{eq: SOE for G}, which involves $\mathcal{O}(N^3)$ computational cost to obtain the non-negative tuples $\{(\omega_j,\lambda_j)\}_{j=1}^{N}$.

\begin{figure}[hbt!]
    \centering
    \vspace{-0.35cm}
    \subfigtopskip = 2pt
    \subfigbottomskip = 2pt
    \subfigcapskip = -2pt
    \subfigure{
    \includegraphics[width=0.45\linewidth,trim={10cm 0.3cm 10cm 0cm},clip]{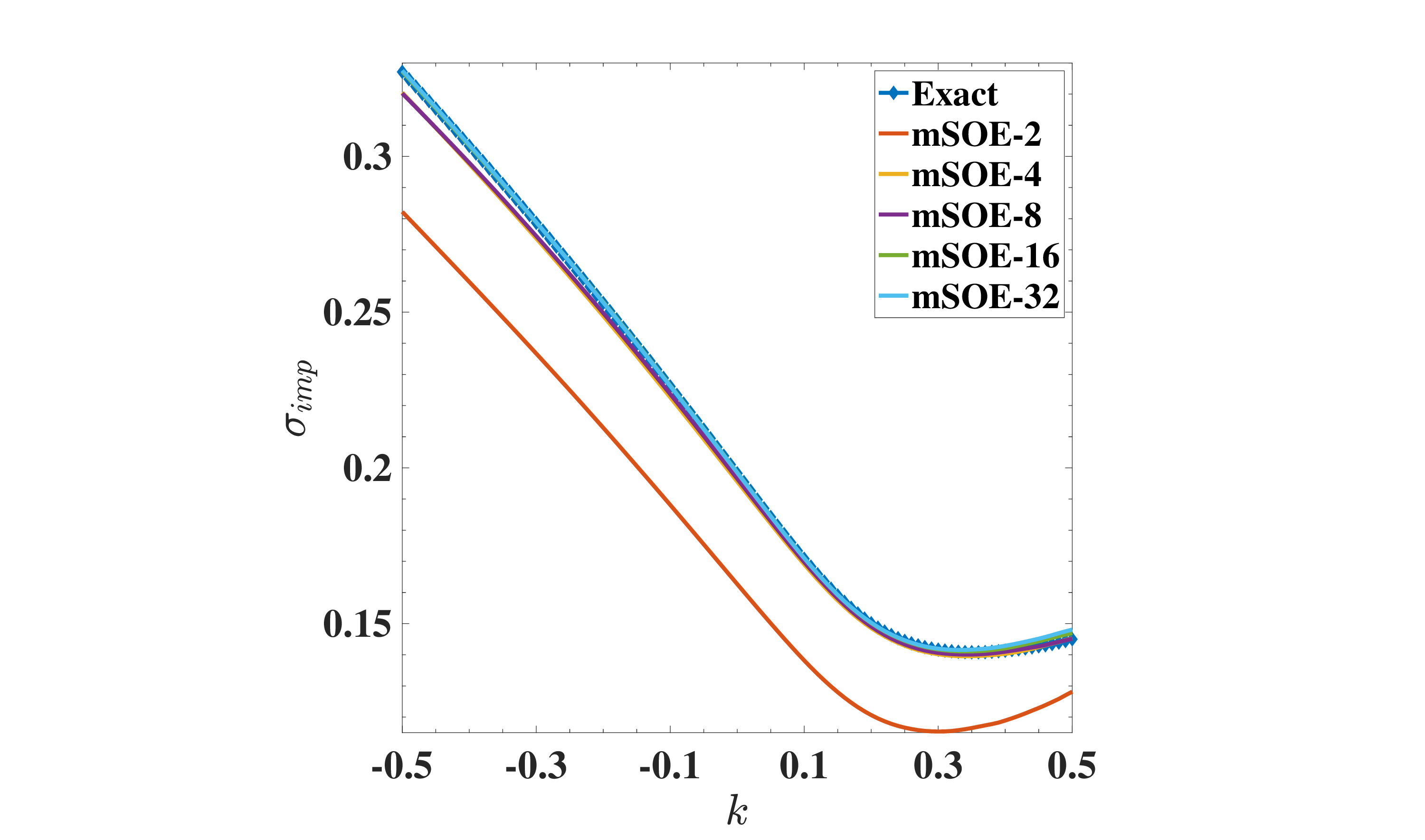}}
    \subfigure{
    \includegraphics[width=0.45\linewidth,trim={10cm 0.3cm 10cm 0cm},clip]{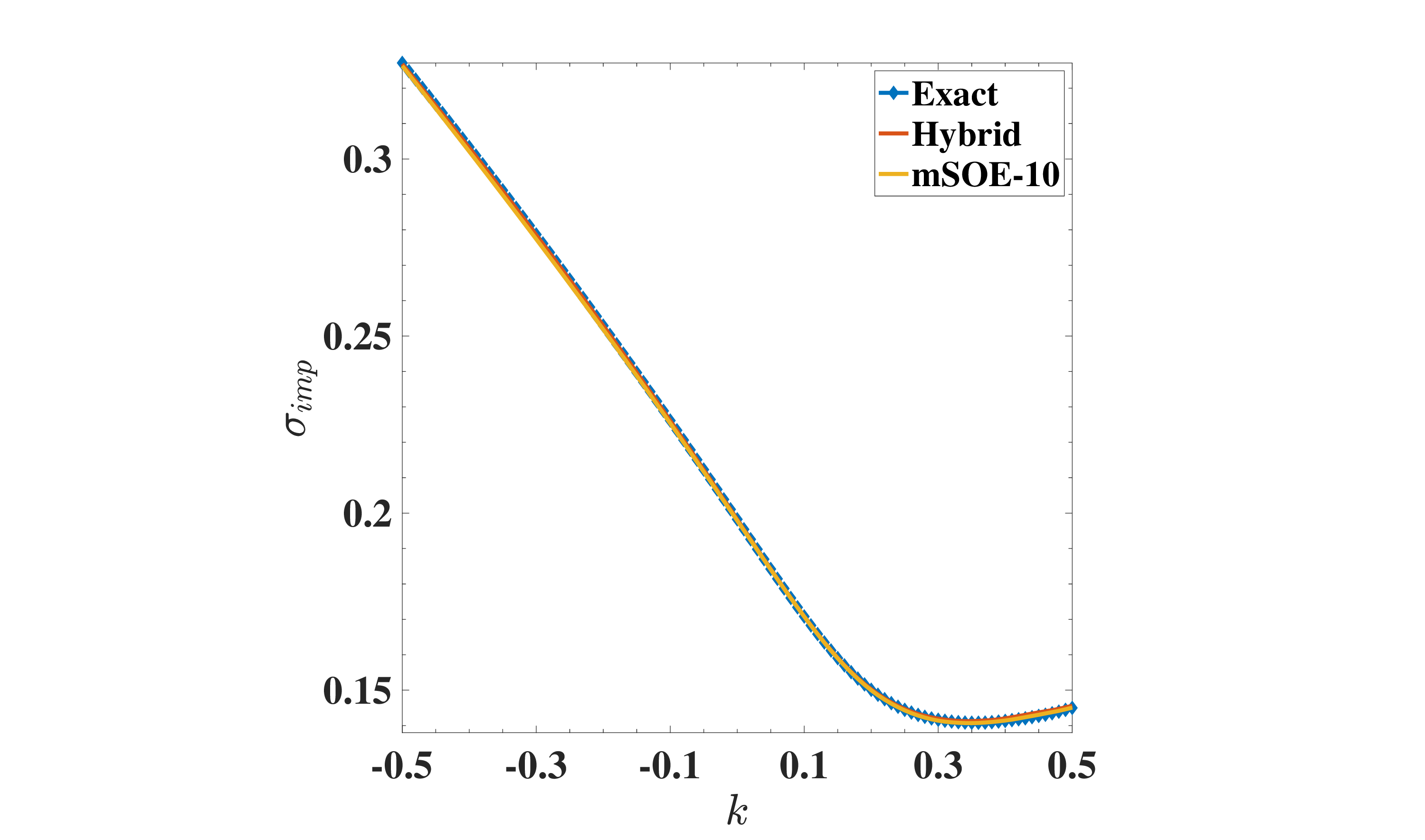}}
    \\
    \caption{The implied volatility curves $\sigma_{imp}$ computed by the mSOE-$N$ based numerical scheme \eqref{eq: SOE for rbergomi} with different number of summation terms $N$. In the plot, $k=\log(K/S_0)$ is the log-moneyness and $K$ is the strike price.}
    \label{Figure: SOE imp vol}
\end{figure}

Next we calculate the implied volatility curves using the mSOE-$N$ based numerical schemes with $N=2,4,8,10,32$ using the parameters listed in Table \ref{Table: params for rBergomi}. To ensure that the temporal discretization error and the Monte Carlo simulation error are negligible, we take the number of temporal steps $n=2000$ and the number of samples $M=10^6$. The resulting implied volatility curves are shown in Figure \ref{Figure: SOE imp vol}, which are solved via the Newton-Raphson method. We compare the proposed mSOE scheme with the exact method, i.e., Cholesky decomposition, and the hybrid scheme. We observe that mSOE-4 scheme can already accurately approximate the implied volatility curves given by the exact method. This clearly shows the high efficiency of the proposed mSOE based scheme for simulating the implied volatility curves.

\begin{table}[!htbp]
    \centering
\begin{threeparttable}
    \caption{Parameter values used in the rBergomi model.} \label{Table: params for rBergomi}
    \begin{tabular}{ccccc}
    \toprule
     $S_0$ & $\xi_0$  & $\eta$ & $H$ & $\rho$ \\
     \midrule
     1 & $0.235^2$ & 1.9 & 0.07 & -0.9 \\
     \bottomrule
    \end{tabular}
\end{threeparttable}
\end{table}

\begin{figure}[hbt!]
    \centering
    \vspace{-0.35cm}
    \subfigtopskip = 2pt
    \subfigbottomskip = -2pt
    \subfigcapskip = -4pt
    \subfigure{
    \includegraphics[width=0.49\linewidth,trim={8cm 0cm 8cm 0cm},clip]{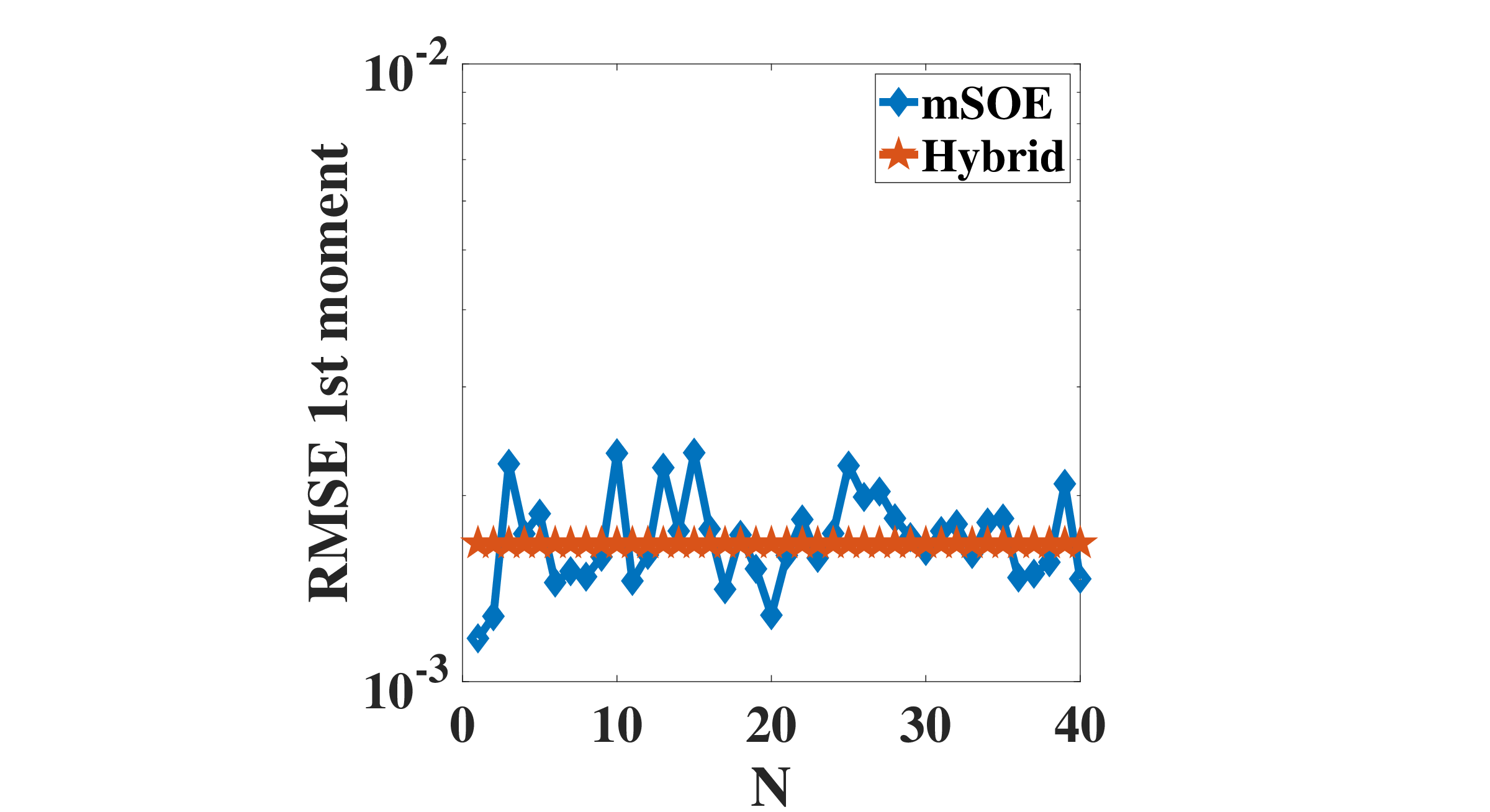}}
    \subfigure{
    \includegraphics[width=0.48\linewidth,trim={8cm 0cm 8cm 0cm},clip]{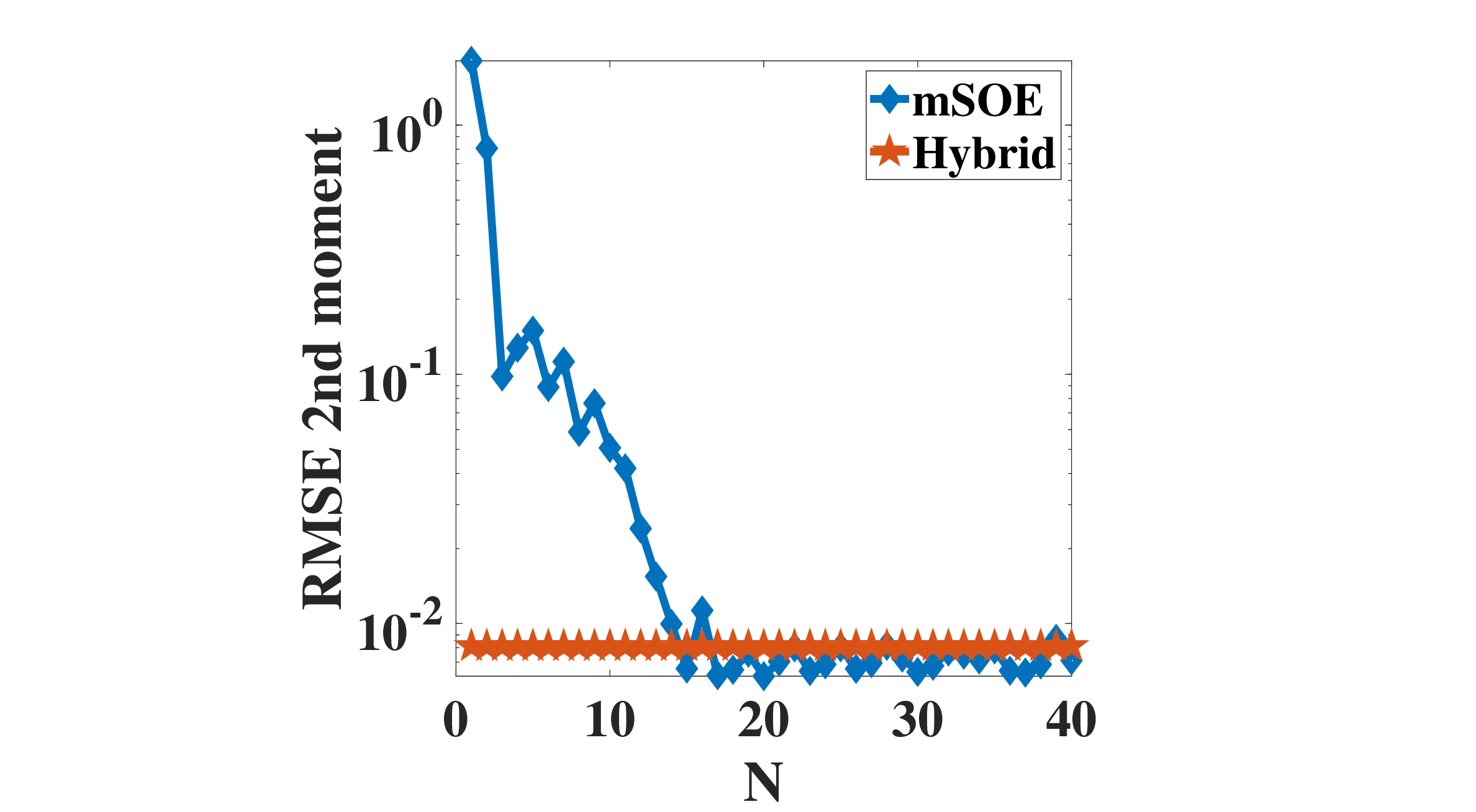}}
    \caption{The RMSEs for the first and second moments generated by the mSOE-$N$ based numerical scheme \eqref{eq: SOE for rbergomi} and the hybrid scheme.}
\label{fig:gaussian-rv}
\end{figure}

Next, we depict in Figure \ref{fig:gaussian-rv} the root mean squared error (RMSE) of the first moment and second moment of the following Gaussian random variable using the same set of parameters,
\begin{align*}
    \mathcal{G}(t) := \eta I_t - \frac{\eta^2}{2}t^{2H} \sim \mathcal{N}\left(-\frac{\eta^2}{2}t^{2H}, \eta^2 t^{2H}\right).
\end{align*}
The RMSEs are defined by
\begin{align*}
\text{RMSE 1st moment} &=\br{ \sum_{i = 1}^n \br{\mathbb{E}\fa{\hat{\mathcal{G}}(t_i)}  + \frac{\eta^2}{2}t_i^{2H}}^2}^{1/2}, \\
\text{RMSE 2nd moment} &= \br{\sum_{i=1}^n\br{\mathbb{E}\fa{\hat{\mathcal{G}}(t_i)^2} - \br{\frac{\eta^4t_i^{4H}}{4} + \eta^2t_i^{2H}}}^2}^{1/2},
\end{align*}
where $\hat{\mathcal{G}}(t)$ is the approximation of $\mathcal{G}(t)$ under the mSOE-N scheme or the hybrid scheme. These quantities characterize the fundamental statistical feature of the distribution. We observe that the mSOE-$N$ based numerical scheme achieves high accuracy for the number of terms $N>15$, again clearly showing the high efficiency of the proposed scheme.

\subsection{Numerical performance for learning the forward variance curve}
Now we validate \eqref{eq: pricing error upper bound} using three examples of ground truth forward variance curves:
\begin{subequations}
\begin{align}
&\xi_0(t)\equiv 0.235^2,  \label{eq: forward_var 1}\\
&\xi_0(t)= 2|W_t|, \label{eq: forward_var 2}\\ 
&\xi_0(t)= 0.1|W^H_t| \text{ with }H=0.07. \label{eq: forward_var 3}
\end{align}
\end{subequations}
The first example \eqref{eq: forward_var 1} corresponds to a constant forward variance curve. The second example \eqref{eq: forward_var 2} takes a scaled sample path of Brownian motion as the forward variance curve, and the third example \eqref{eq: forward_var 3} utilises a scaled path of fractional Brownian motion as the forward variance curve.

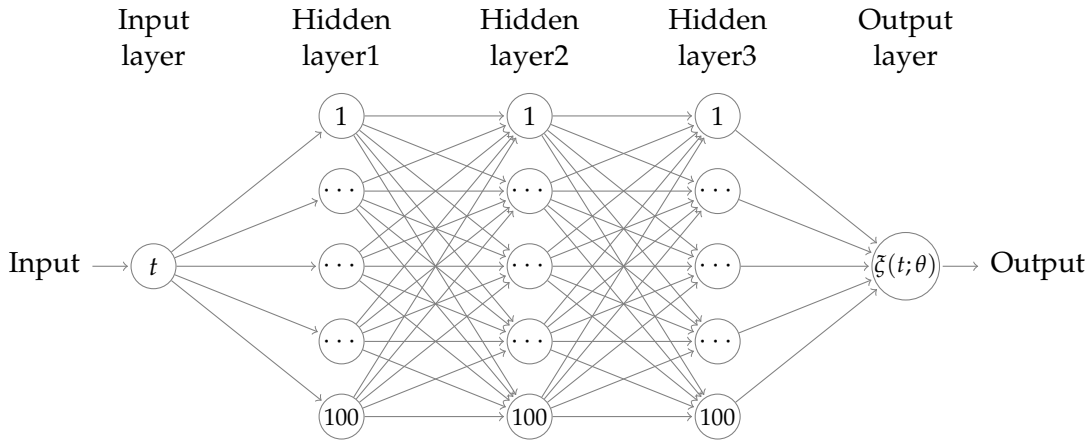
\begin{figure}
\def\layersep{2.5cm}
\centering
\begin{tikzpicture}[shorten >=1pt,->,draw=black!50, node distance=\layersep]
    \tikzstyle{every pin edge}=[<-,shorten <=1pt]
    \tikzstyle{neuron}=[circle,draw,fill=black!50,minimum size=17pt,inner sep=0pt]
    \tikzstyle{input neuron}=[neuron, fill=black!0];
    \tikzstyle{output neuron}=[neuron, fill=black!0];

    \tikzstyle{hidden neuron1}=[neuron, fill=black!0];
    \tikzstyle{hidden neuron2}=[neuron, fill=black!0];
    \tikzstyle{hidden neuron3}=[neuron, fill=black!0];

    \tikzstyle{annot} = [text width=4em, text centered]

    \node[input neuron, pin=left:Input] (I-1) at (0,-3) {\small$t$};

    \node[hidden neuron1] (H1-1) at (\layersep,-1 cm) {\small$1$};
    \foreach \y in {2,3,4}
        \node[hidden neuron1] (H1-\y) at (\layersep,-\y cm) {\small$\cdots$};
    \node[hidden neuron1] (H1-5) at (\layersep,-5 cm) {\footnotesize$100$};

    \node[hidden neuron2] (H2-1) at (2*\layersep,-1 cm) {\small$1$};
    \foreach \y in {2,3,4}
        \node[hidden neuron2] (H2-\y) at (2*\layersep,-\y cm) {\small$\cdots$};
    \node[hidden neuron2] (H2-5) at (2*\layersep,-5 cm) {\footnotesize$100$};

    \node[hidden neuron3] (H3-1) at (3*\layersep,-1 cm) {\small$1$};
    \foreach \y in {2,3,4}
        \node[hidden neuron3] (H3-\y) at (3*\layersep,-\y cm) {\small$\cdots$};
    \node[hidden neuron3] (H3-5) at (3*\layersep,-5 cm) {\footnotesize$100$};
    \node[output neuron,pin={[pin edge={->}]right:Output}, right of=H3-3] (O) {\footnotesize$\xi(t;\theta)$};

    \foreach \source in {1}
        \foreach \dest in {1,...,5}
            \path (I-\source) edge (H1-\dest);

    \foreach \source in {1,...,5}
        \foreach \dest in {1,...,5}
            \path (H1-\source) edge (H2-\dest);

    \foreach \source in {1,...,5}
        \foreach \dest in {1,...,5}
            \path (H2-\source) edge (H3-\dest);
    \foreach \source in {1,...,5}
        \path (H3-\source) edge (O);
    \node[annot,above of=H1-1, node distance=1cm] (hl) {Hidden layer1};
    \node[annot,left of=hl] {Input layer};
    \node[annot,right of=hl] (h2){Hidden layer2};
    \node[annot,right of=h2] (h3){Hidden layer3};
    \node[annot,right of=h3] {Output layer};
\end{tikzpicture}
\caption{Architecture: 3-layer feedforward neural network.}
\label{fig:architecture}
\end{figure}

The initial forward variance curve $\xi_0$ is parameterized by a feed forward neural network, cf. \eqref{eq:neural-variance}, which has 3 hidden layers, width 100, and leaky ReLU activations; see Fig. \ref{fig:architecture} for  a schematic illustration. The weights of each model were carefully initialized in order to prevent gradient vanishing.

To generate the training data, we use the stock price samples generated by the scheme \eqref{eq: SOE for rbergomi}. The total number of samples used in the experiments is $10^5$. Training of the neural network was performed on the first $81.92\%$ of the dataset and the model's performance was evaluated on the remaining $18.08\%$ of the dataset. Each sample is of length 2000, which is the number of discretized time intervals. The batch size was 4096, which was picked as large as possible that the GPU memory allowed for. Each model was trained for 100 epochs. We consider Wasserstein-1 distance as loss metric by applying \eqref{eq: empirical W_p} to the empirical distributions on batches of data at the final time grid of the samples.

The neural SDEs were trained using Adam \cite{KingmaBa:2015}. The learning rate was taken to be $10^{-4}$, which was then gradually reduced until a good performance was achieved. The training was performed on a Linux server and a NVIDIA RTX6000 GPU and takes a few hours for each experiment. $V_t(\theta)$ in \eqref{eq: rbergomi_v-nn} was solved by the mSOE scheme to reduce the computational complexity, then $S_t(\theta)$ was solved using the Euler-Maruyama method. The example code for dataset sampling and network training will be made available at the Github repository {\tiny\url{https://github.com/evergreen1002/Neural-option-pricing-for-rBergomi-model}}.

\begin{figure}[H]
    \centering
    \vspace{-0.35cm}
    \subfigtopskip = 2pt
    \subfigbottomskip = 2pt
    \subfigcapskip = -2pt
    \subfigure{
    \includegraphics[width=0.18\linewidth,trim={12cm 0.5cm 13cm 1cm},clip]{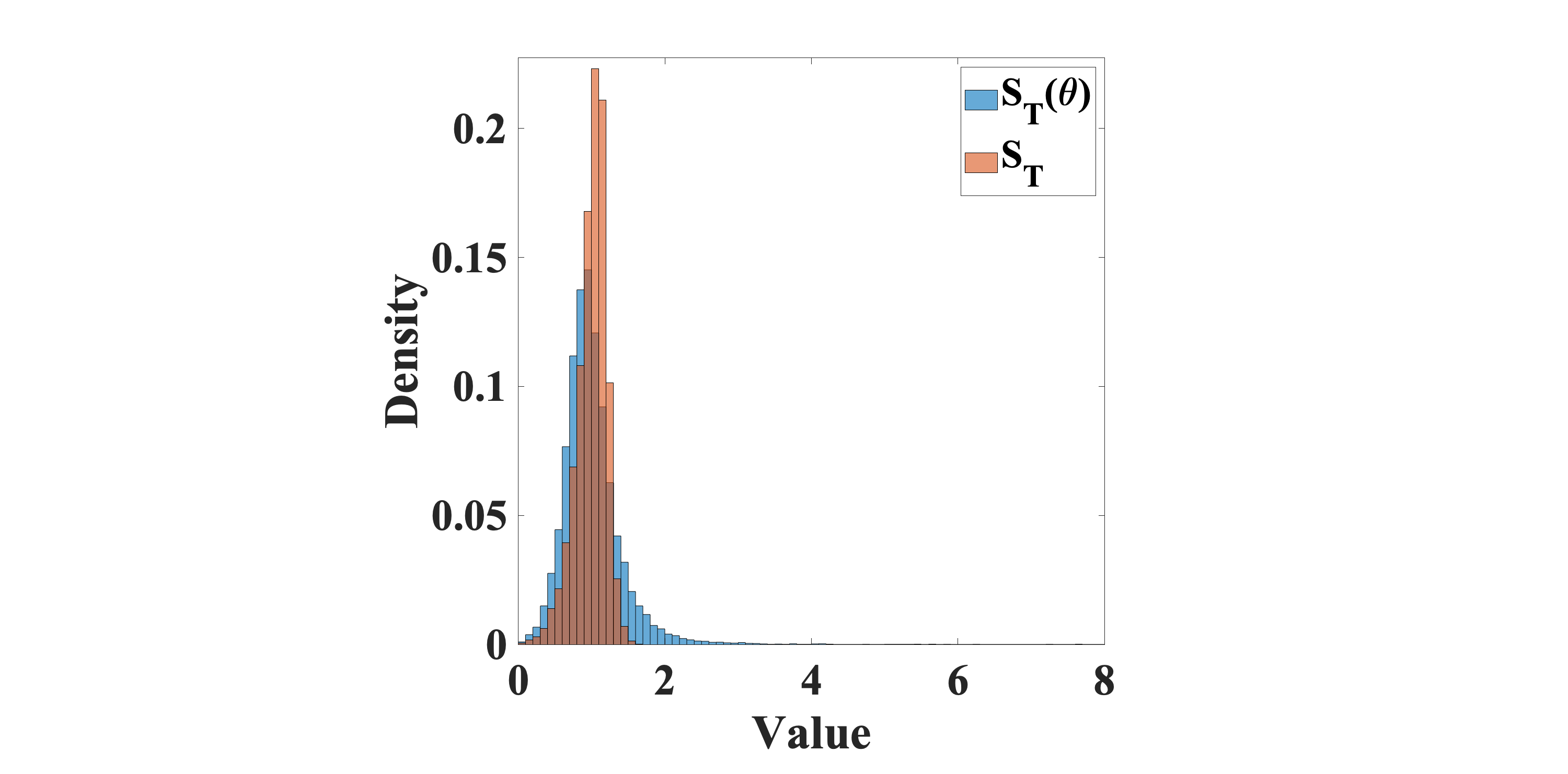}}
 \subfigure{
    \includegraphics[width=0.18\linewidth,trim={12cm 0.5cm 13cm 1cm},clip]{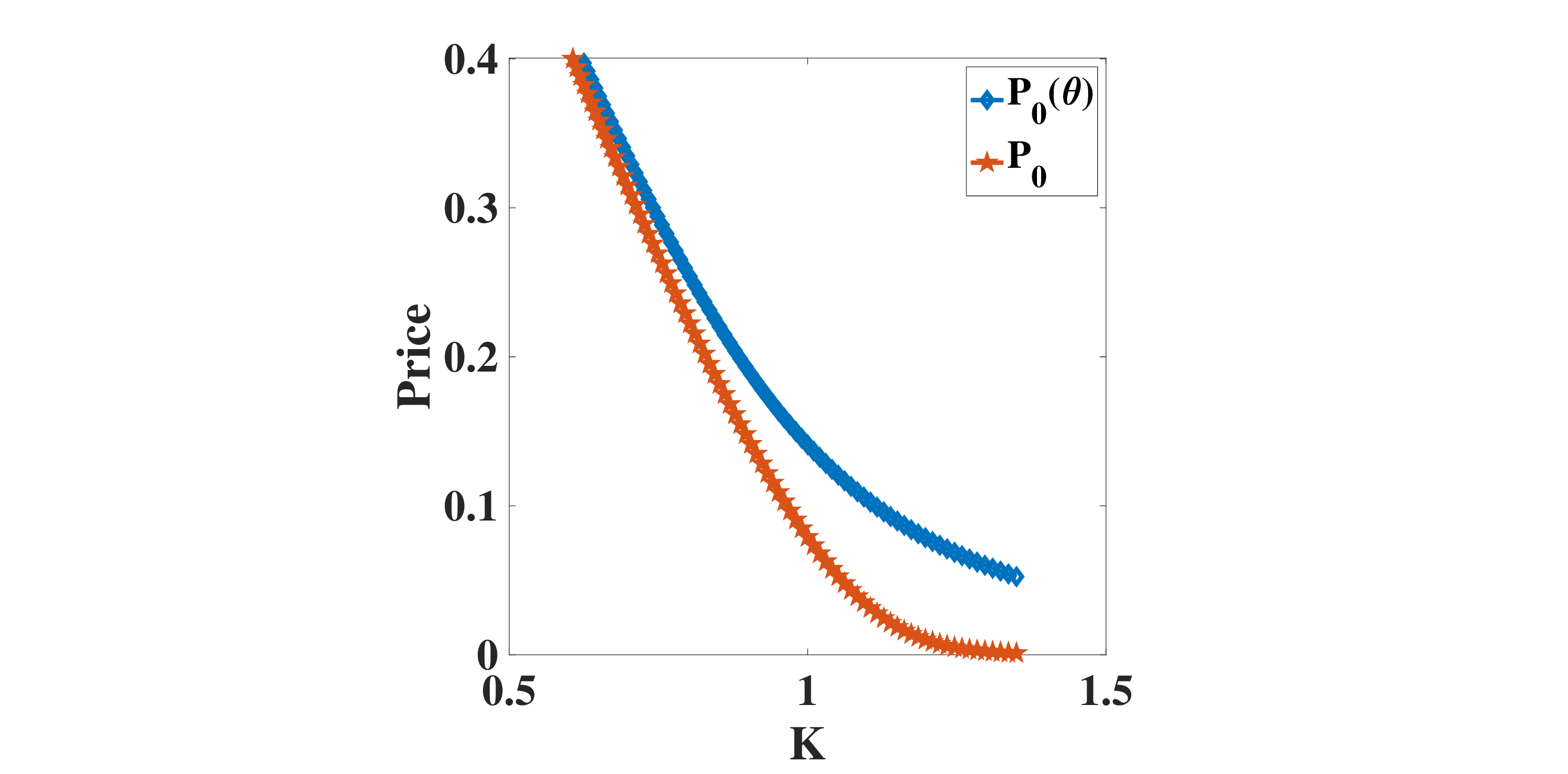}}
    \subfigure{
    \includegraphics[width=0.18\linewidth,trim={12cm 0.5cm 13cm 1cm},clip]{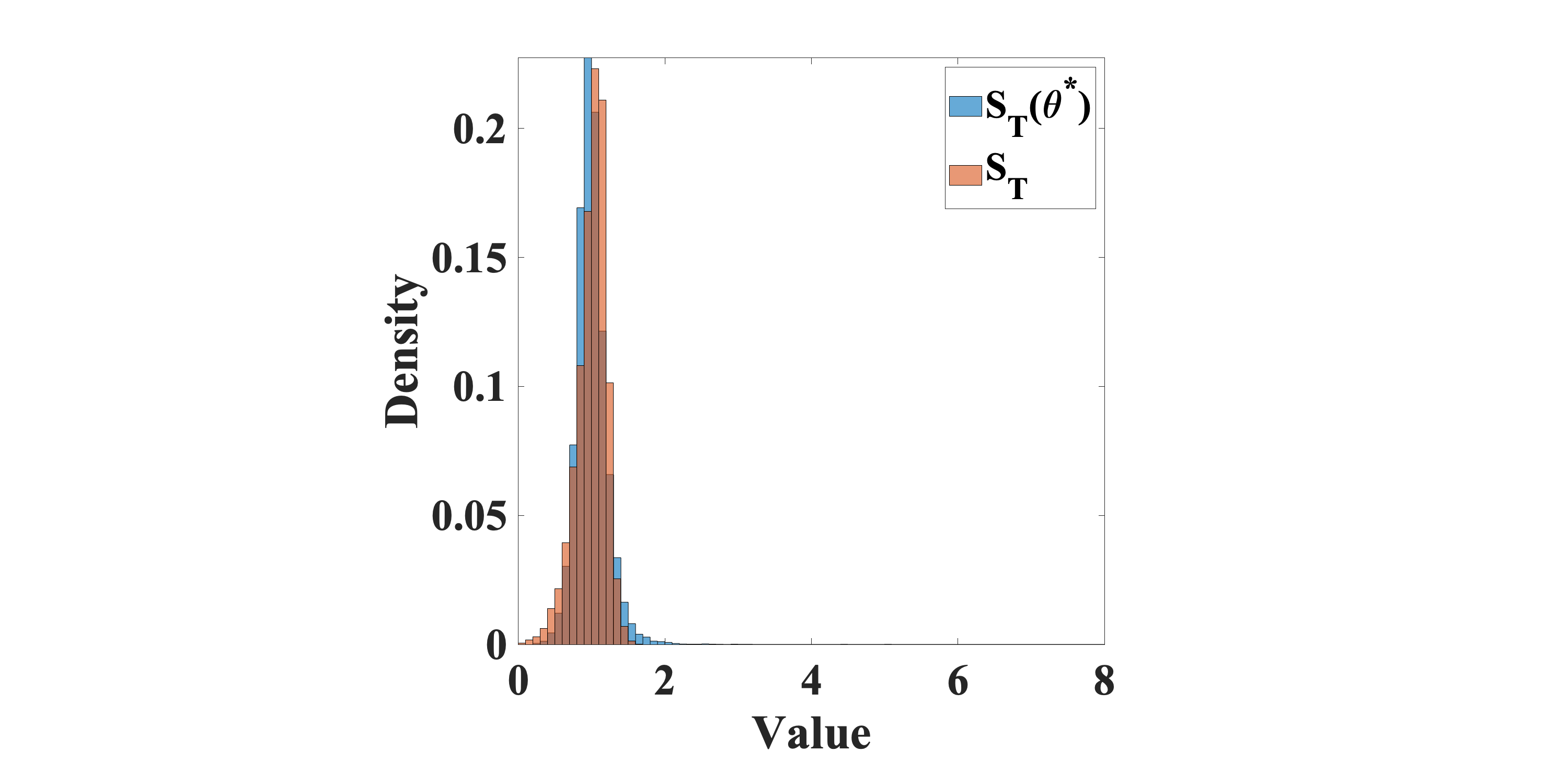}}
    \subfigure{
     \includegraphics[width=0.18\linewidth,trim={14cm 0.2cm 14cm 1cm},clip]{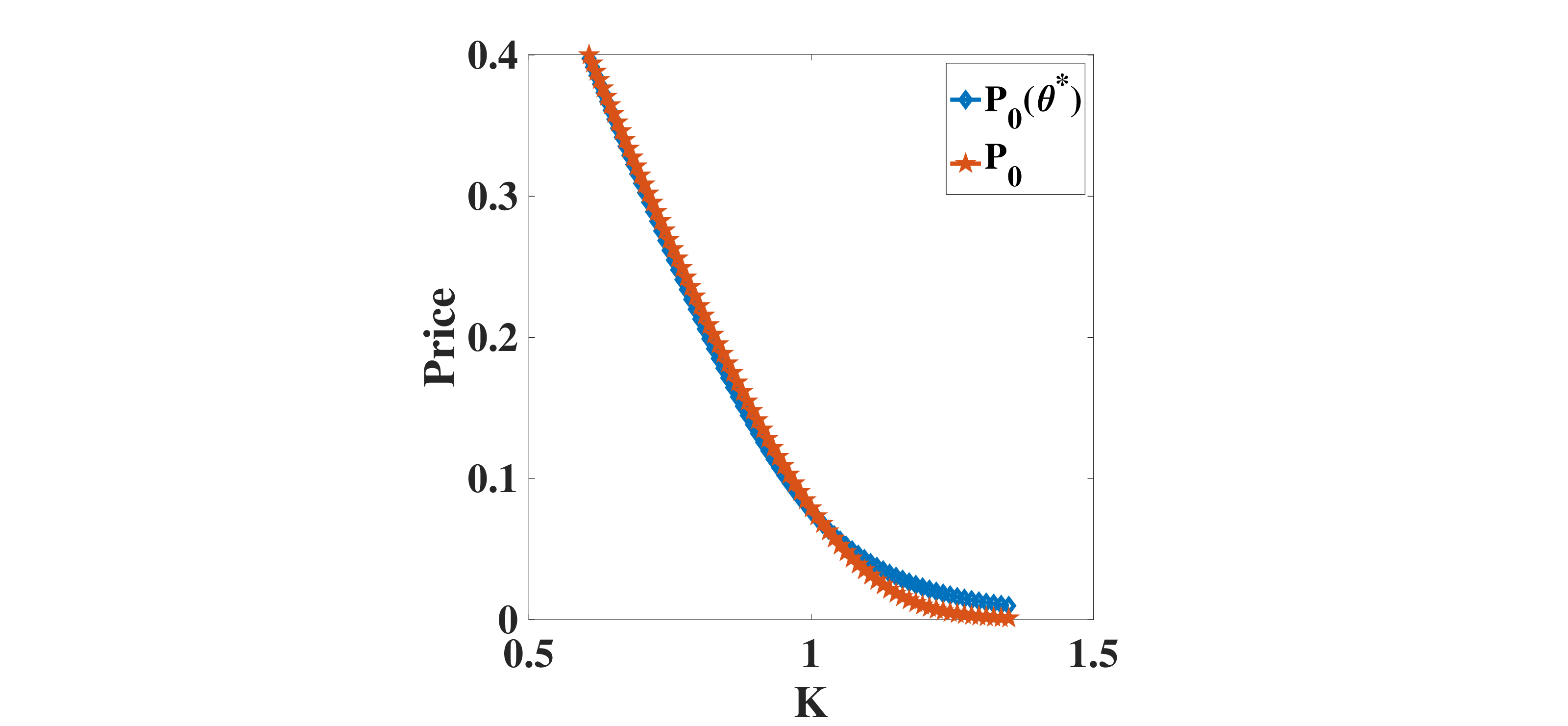}}
 \subfigure{
     \includegraphics[width=0.185\linewidth,trim={12cm 0.5cm 12cm 1cm},clip]{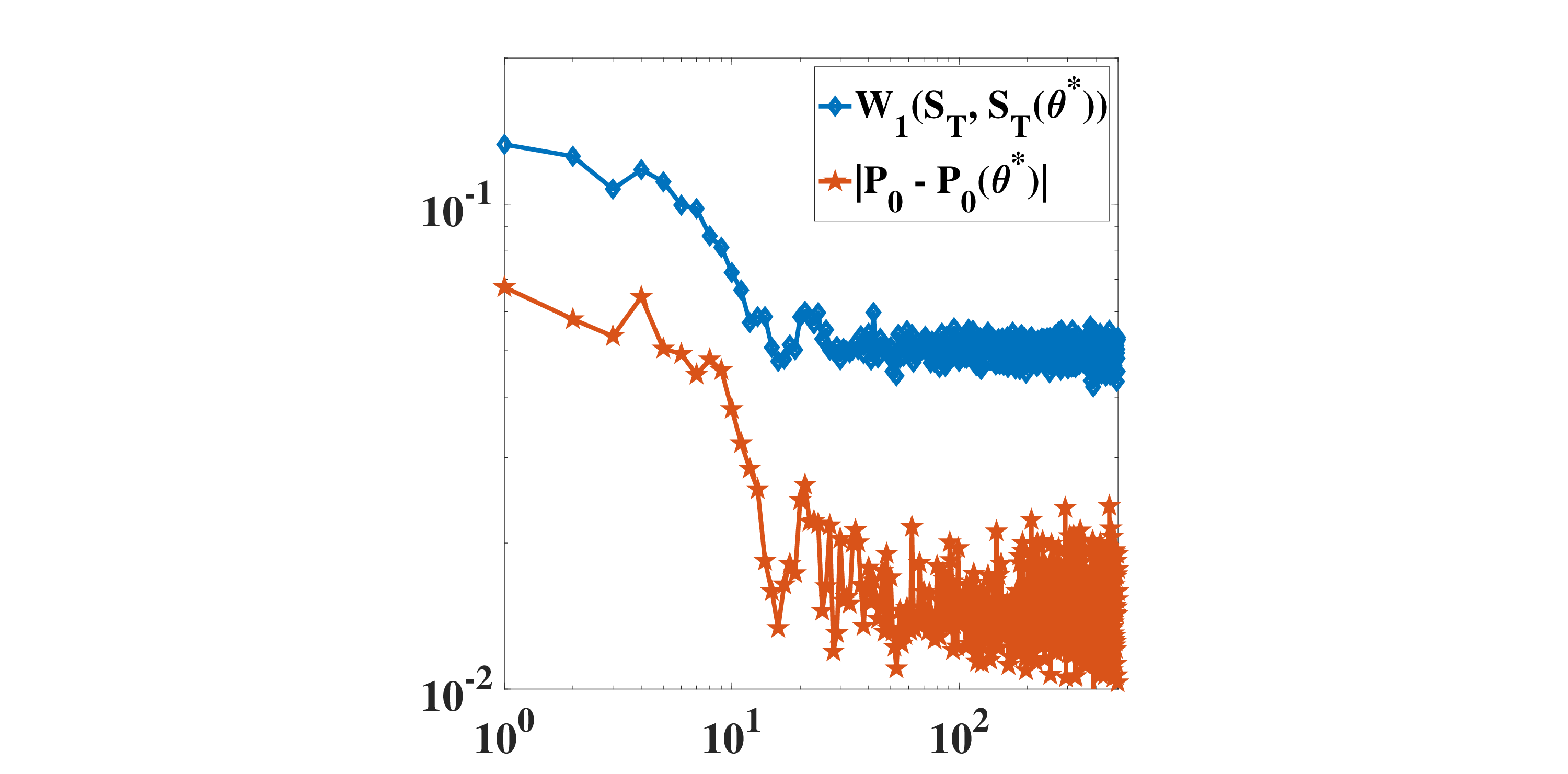}}
    \\

    \subfigure{
    \includegraphics[width=0.18\linewidth,trim={12cm 0.5cm 13cm 1cm},clip]{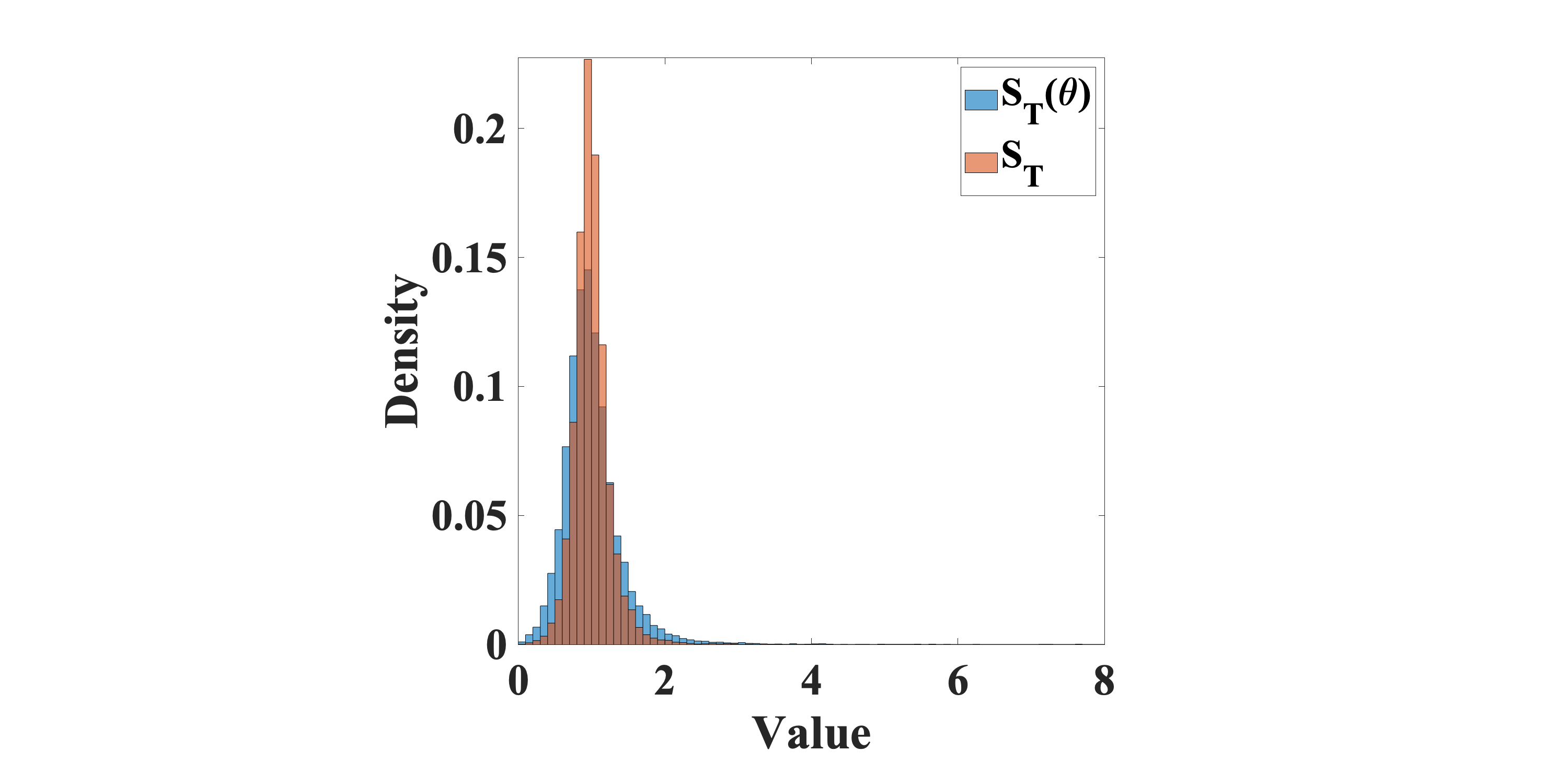}}
 \subfigure{
    \includegraphics[width=0.18\linewidth,trim={14cm 0.2cm 14cm 1cm},clip]{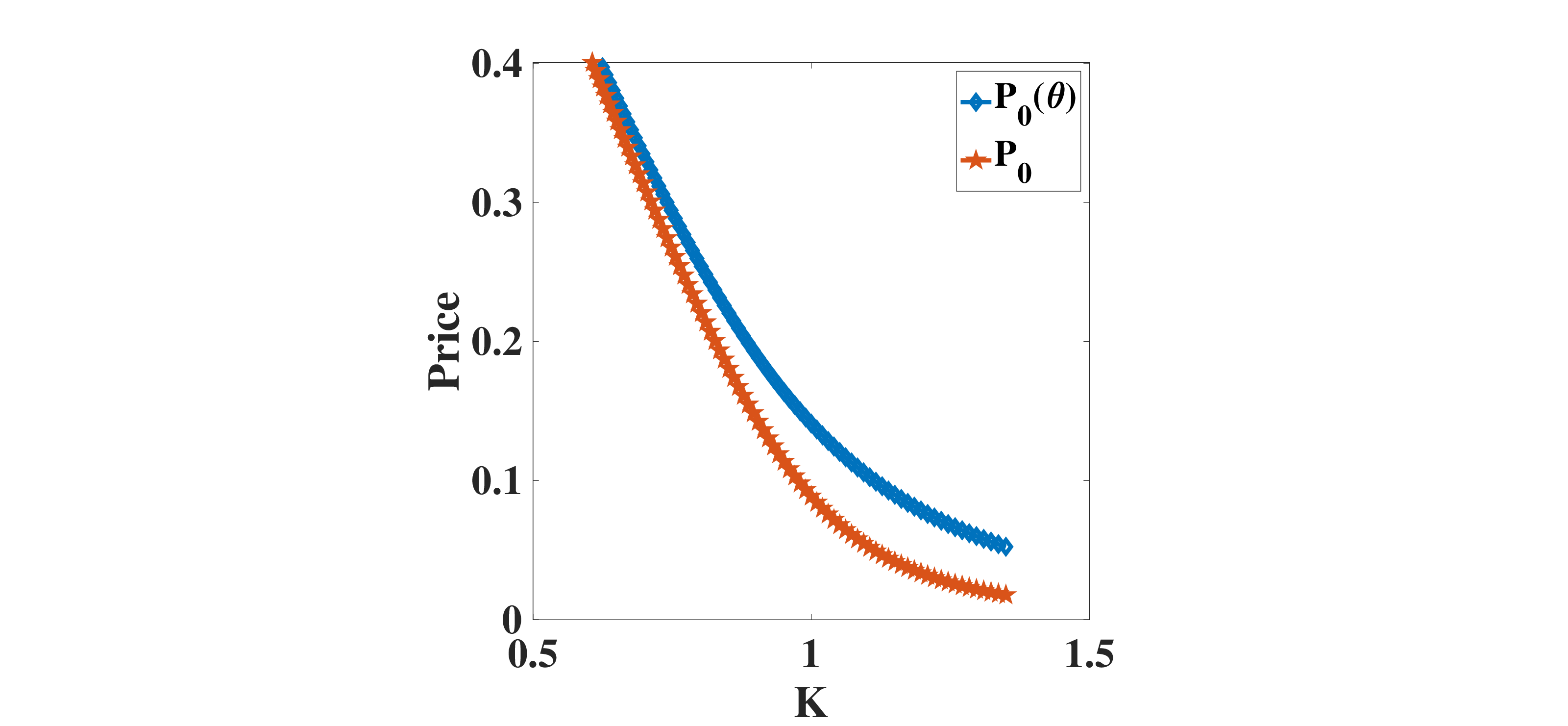}}
    \subfigure{
    \includegraphics[width=0.18\linewidth,trim={12cm 0.5cm 13cm 1cm},clip]{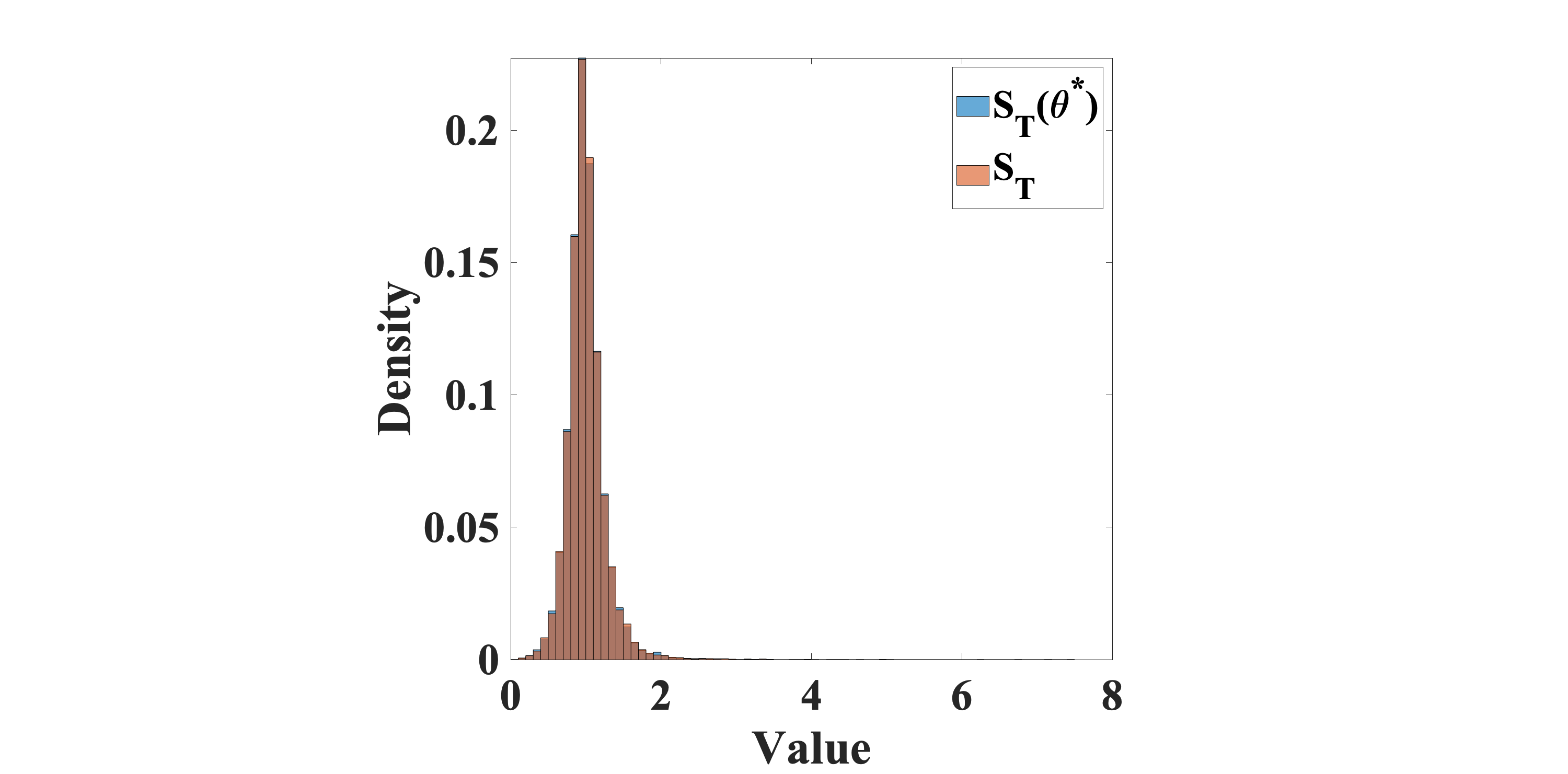}}
    \subfigure{
     \includegraphics[width=0.18\linewidth,trim={12cm 0.2cm 12cm 1cm},clip]{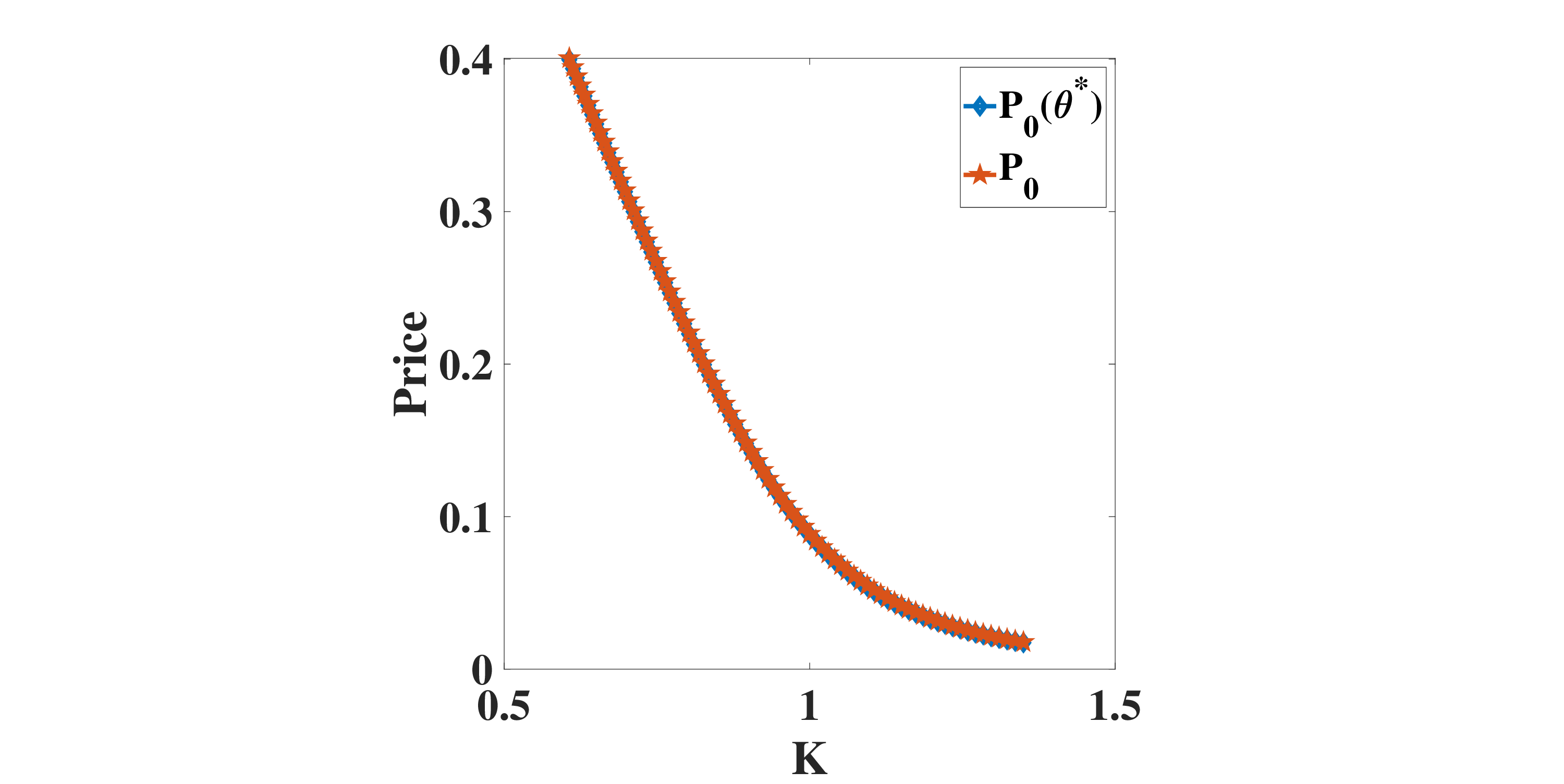}}
 \subfigure{
     \includegraphics[width=0.18\linewidth,trim={11cm 0.5cm 11cm 1cm},clip]{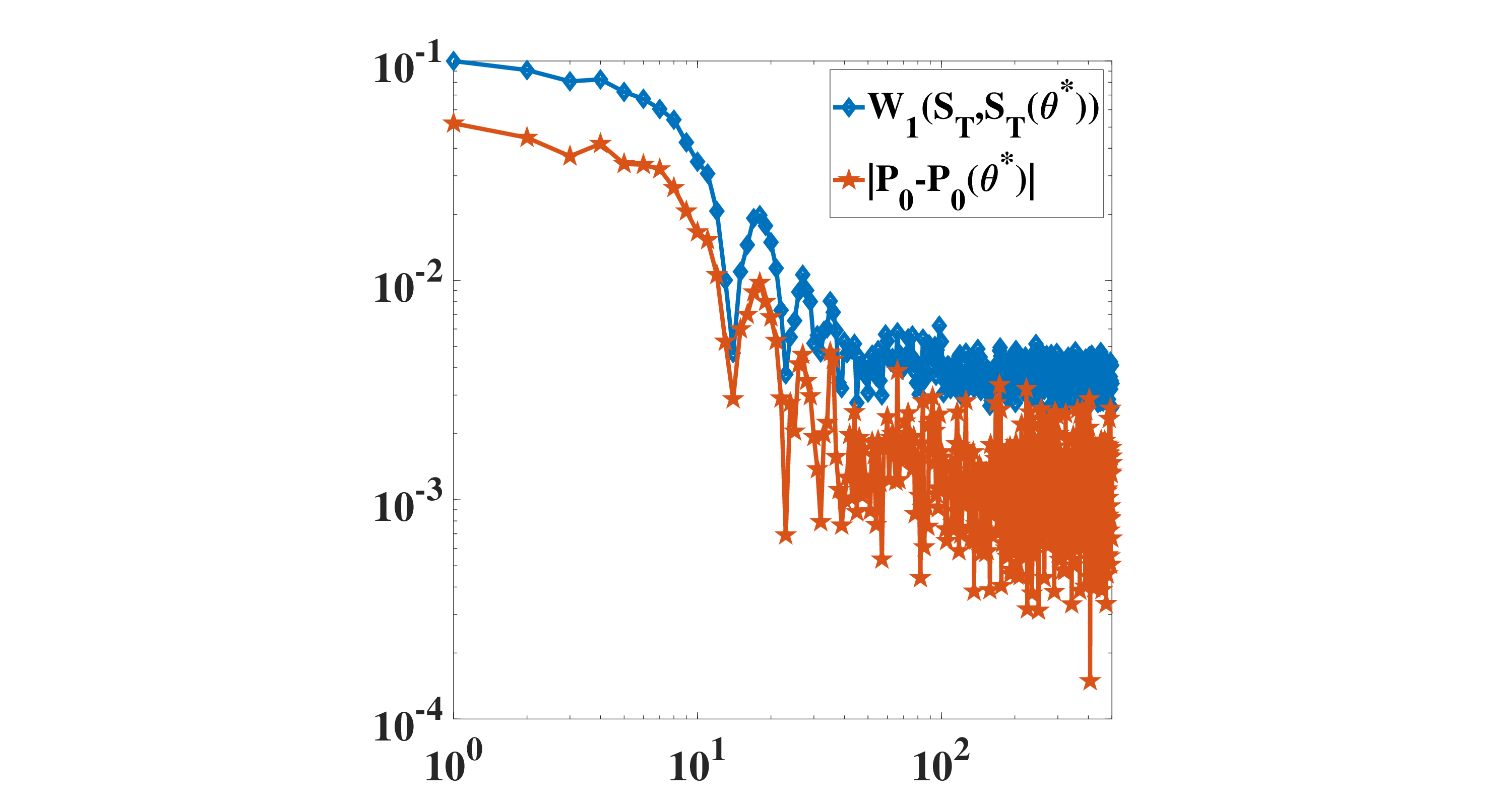}}
    \\

\subfigure{
    \includegraphics[width=0.18\linewidth,trim={12cm 0.5cm 13cm 1cm},clip]{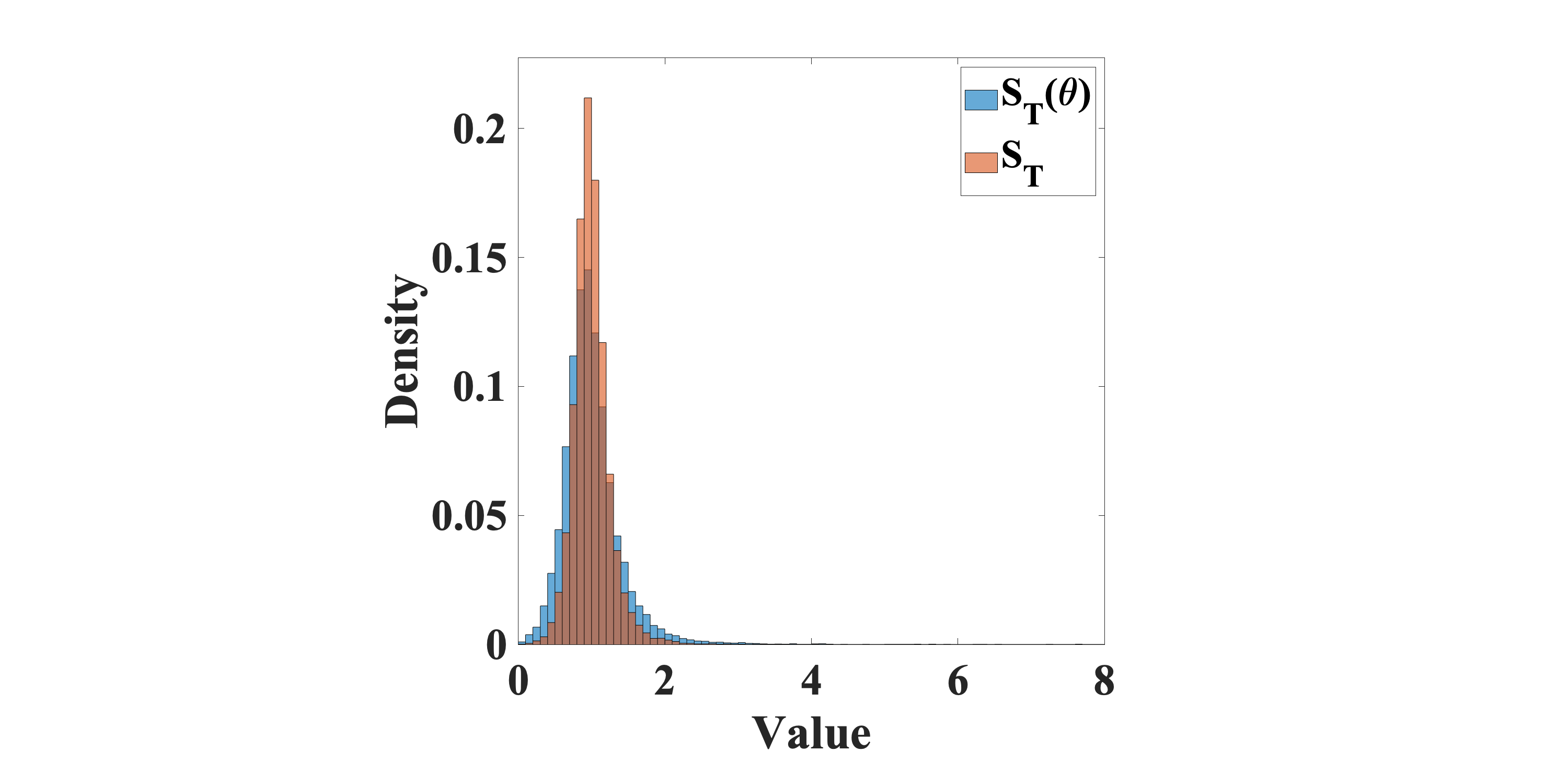}}
 \subfigure{
    \includegraphics[width=0.18\linewidth,trim={14cm 0.2cm 14cm 1cm},clip]{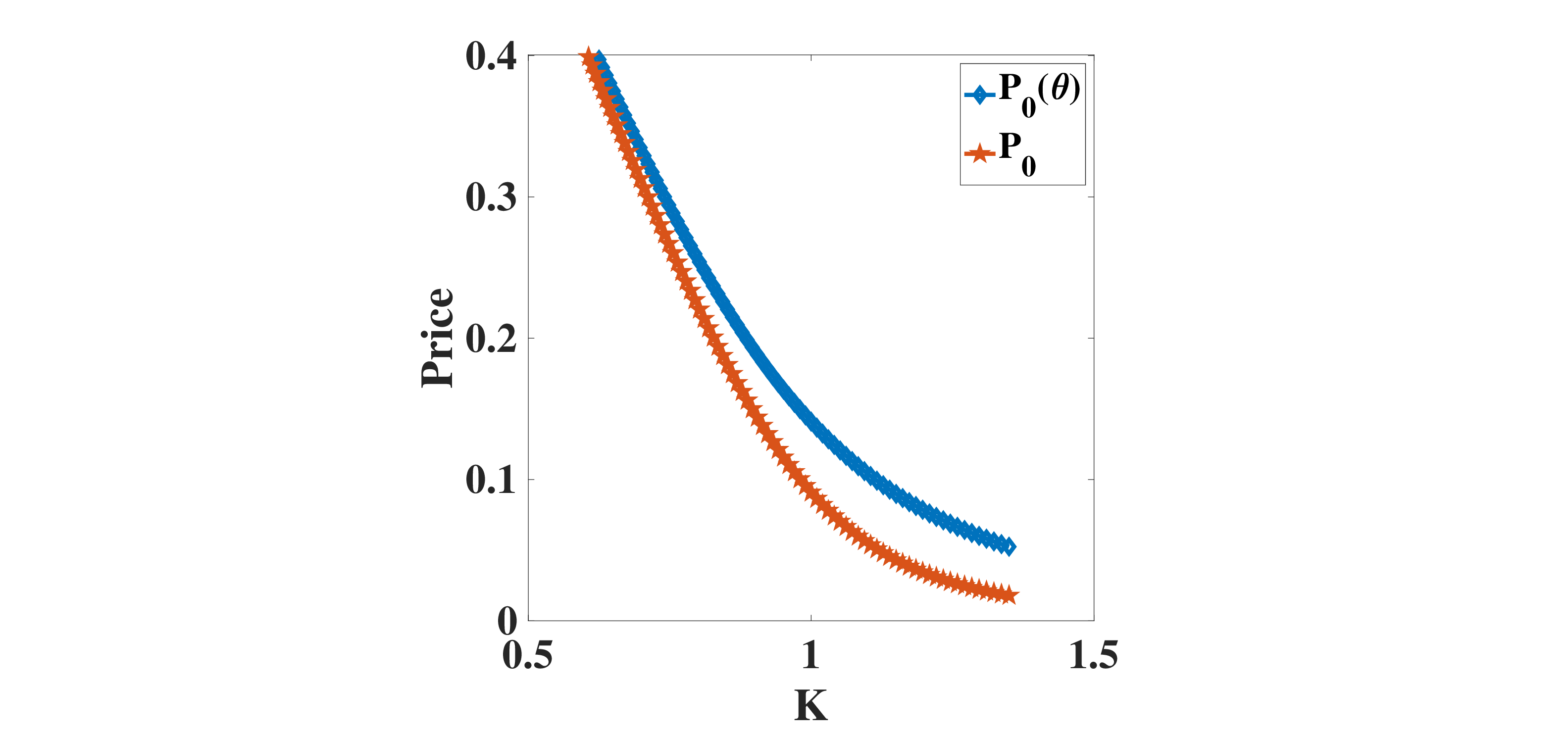}}
    \subfigure{
    \includegraphics[width=0.18\linewidth,trim={12cm 0.5cm 13cm 1cm},clip]{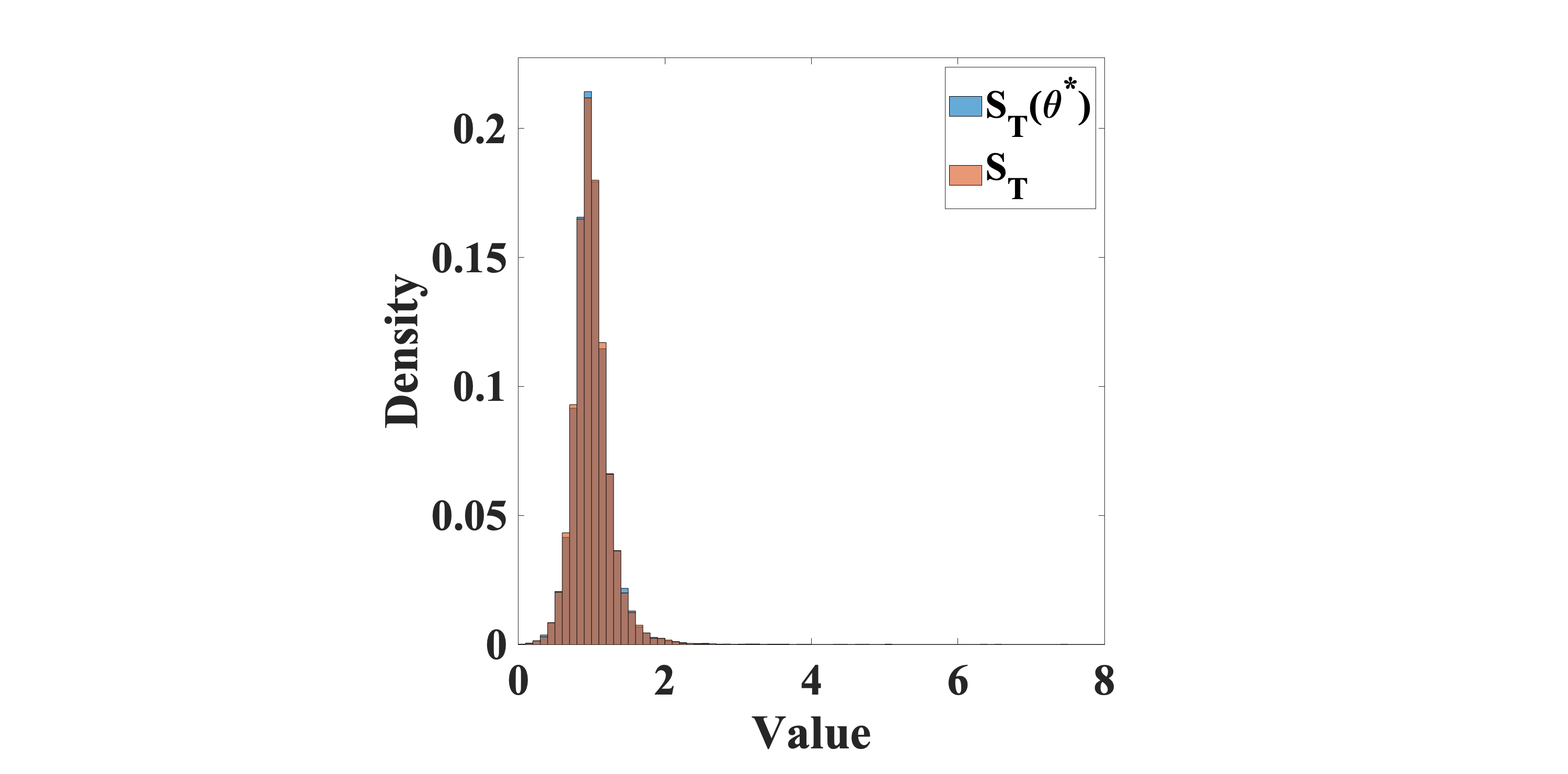}}
    \subfigure{
     \includegraphics[width=0.18\linewidth,trim={14cm 0.2cm 14cm 1cm},clip]{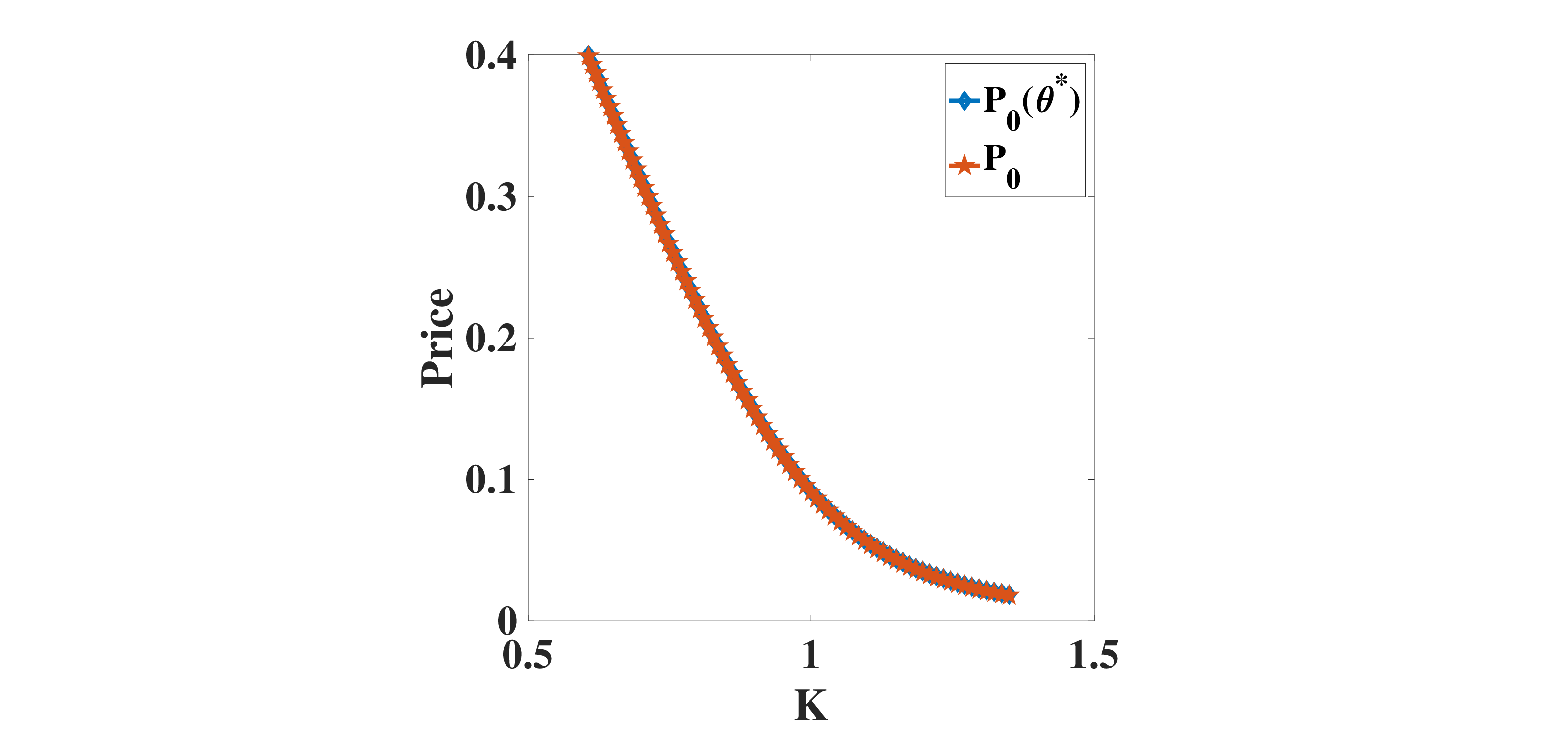}}
 \subfigure{
     \includegraphics[width=0.18\linewidth,trim={12cm 0.5cm 12cm 1cm},clip]{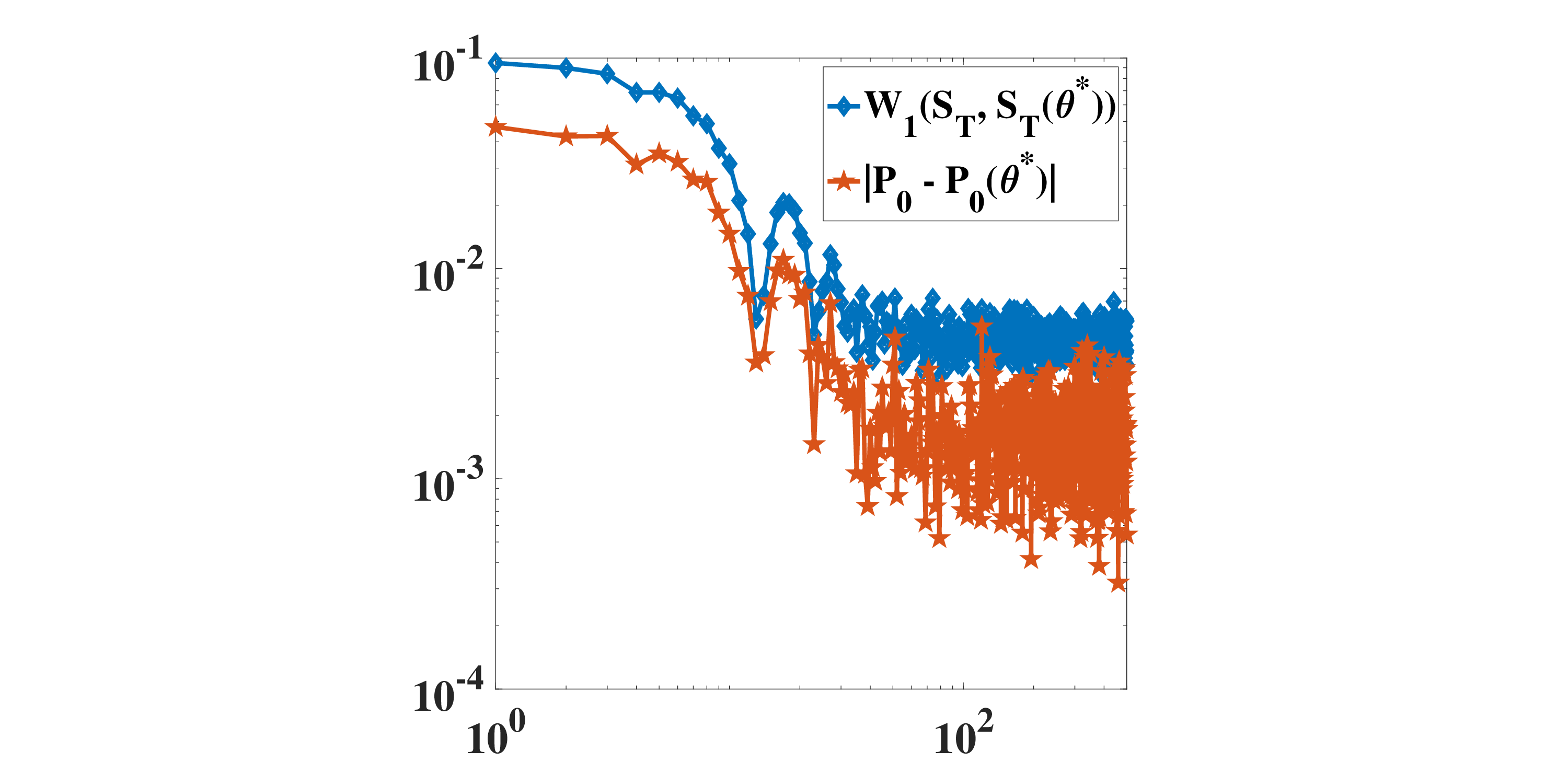}}

    \caption{Left to right: Empirical distribution of terminal stock price before the training, option price before the training, empirical distribution of terminal stock price after the training, option price after training, learning curve.}
    \label{fig:train-results}
\end{figure}

Now we present the numerical results for the learned forward variance curves in Figure \ref{fig:train-results}. These three rows correspond to three different initial forward variance curves defined in \eqref{eq: forward_var 1}, \eqref{eq: forward_var 2} and \eqref{eq: forward_var 3} used for neural network training, respectively.
The first two columns display the empirical distributions $S_T(\theta)$ and the corresponding European call option prices $P_0(\theta)$ for the test set against a number of strikes before training has begun. The next two columns display $S_T(\theta^*)$ and $P_0(\theta^*)$ after training. It is observed that $S_T(\theta^*)$ and $P_0(\theta^*)$ closely match the exact one on the test set, thereby showing the high accuracy of learning the rBergomi model. The last column shows the Wasserstein-1 distance at time $T$ compared to the maximum error in the option price over each batch during training. These two quantities are precisely the subjects in \eqref{eq: pricing error upper bound}. Only the first $500$ iterations were plotted. In all cases, the training loss can be reduced to an acceptable level, so is the error of pricing, clearly indicating the feasibility of learning within the rBergomi model. Finally, we investigate whether the training loss can be further reduced after more iterations for the first example \eqref{eq: forward_var 1}. Figure \ref{fig: kernel approximation}(b) shows the training loss against the number of iterations. Gradient descent is applied with learning rate $10^{-5}$. We observe that the training loss is oscillating near 0.05 even after 100,000 iterations. Indeed, the oscillation of the loss values aggravates as the iteration further proceeds, indicating the necessity for early stopping of the training process.

\section{Conclusion}\label{sec: conclusion}
In this work, we have proposed a novel neural network based SDE model to learn the forward variance curve for the rough Bergomi model. We propose a new modified summation of exponentials (mSOE) scheme to improve the efficiency of generating training data and facilitating the training process. We have utilized the Wasserstein 1-distance as the loss function to calibrate the dynamics of the underlying assets and the price of the European options simultaneously. Furthermore, several numerical experiments are provided to demonstrate their performances, which clearly show the feasibility of neural networks for calibrating these models. Future work includes learning all the unknown functions in the rBergomi model (as well as other rough volatility models) by the proposed approach, using the market data instead of the simulated data as in the present work.

\bibliographystyle{abbrv}
\bibliography{myref}
\end{document}